\documentclass[conference]{IEEEtran}
\IEEEoverridecommandlockouts
\usepackage{cite}
\usepackage{amsmath,amssymb,amsfonts} 
\usepackage{algorithmic}
\usepackage{graphicx}
\usepackage{textcomp}
\usepackage{xcolor}
\def\BibTeX{{\rm B\kern-.05em{\sc i\kern-.025em b}\kern-.08em
    T\kern-.1667em\lower.7ex\hbox{E}\kern-.125emX}}

\usepackage{url}                  
\usepackage{listings}          
        
\usepackage{hyperref}
 
\usepackage[utf8]{inputenc} 
\usepackage[T1]{fontenc} 
\usepackage{amsmath}
\usepackage{bm}
 
\usepackage{fancyvrb}
\usepackage{color}
\usepackage{fancybox}
\usepackage[english]{babel}
\usepackage{float,multirow}
\usepackage{mathpartir}
\usepackage{parskip}
\usepackage{stmaryrd}
\usepackage{etoolbox} 
\usepackage[appendix=inline]{apxproof} 
 
\usepackage{letltxmacro}

\LetLtxMacro\oldttfamily\ttfamily
\DeclareRobustCommand{\ttfamily}{\oldttfamily\csname ttsize\endcsname}
\newcommand{\setttsize}[1]{\def\ttsize{#1}}%

\newtheoremrep{defi}{Definition}
\newtheoremrep{prop}{Proposition}
\newtheoremrep{lem}{Lemma} 
\newtheorem{ex}{Example}

\usepackage{stackengine}

\usepackage{graphicx}
\usepackage{algorithm}
\usepackage[scaled]{helvet} 
\usepackage{url}                  
\usepackage{listings}          
\usepackage{subcaption} 
\usepackage{xcolor}
\definecolor{darkgreen}{rgb}{0,0.45,0.08} 
\definecolor{darkblue}{rgb}{0.0, 0.0, 0.55} 
\lstdefinelanguage{XQuery}{
        sensitive=true,
        alsodigit={-},
        morekeywords=[1]{ancestor,ancestor-or-self,child,descendant,descendant-or-self,following,following-sibling,parent,preceding,preceding-sibling,self},
        otherkeywords={/,::,*,[,]},
        morekeywords=[2]{let,for,in,return,if,then,else, while},
        moredelim=[s][\normalsize\ttfamily\color{darkgreen}]{<}{>}
}
\newenvironment{packeditemize}{
\begin{itemize}
  \setlength{\itemsep}{1pt}
  \setlength{\parskip}{0pt}
  \setlength{\parsep}{0pt}
}{\end{itemize}}

\begin{document}\sloppy\newcommand{\mus}{{$\mu$-RA}}
\newcommand{\muir}{{$\mu$IR}}
\newcommand{\distmuir}{Dist-$\mu$-RA}
\newcommand{\todo}[1]{\textcolor{blue}{\textsc{todo:}{ #1}}}

\newcommand{\tcblue}[1]{\color{darkblue}{#1} \color{black}}
\newcommand{\removerev}[1]{\color{darkgreen}{#1}\color{black}}
\newcommand{\AJoin}{\triangleright}
\newcommand{\NJoin}{\bowtie}
\newcommand{\fixpt}[2][X]{\ensuremath{\mu(#1=#2)}}
\newcommand{\sem}[2]{{\llbracket#1\rrbracket}_{#2}}
\newcommand{\trad}[2]{{{#1}(#2)}}
\newcommand{\dom}[1]{dom(#1)}
\newcommand{\rename}[3]{\rho_{#1}^{#2}\left(#3\right)}
\newcommand{\drop}[2]{\ensuremath{\widetilde{\pi}_{#1}(#2)}}
\newcommand{\proj}[2]{\pi_{#1}\left(#2\right)}
\newcommand{\filt}[2][\mathfrak f]{\ensuremath{\sigma_{#1}\left(#2\right)}}
\newcommand{\cst}{c\rightarrow v}
\newcommand{\agg}[2][]{\Theta\left(#2,g{#1},{\C}{#1}, {\D}{#1}\right) }
\newcommand{\specialagg}[4]{\Theta\left(#4,#1,{#2}, {#3}\right) }

\newcommand{\filtname}[1][\mathfrak f]{\sigma_{#1}}

\newcommand{\mapping}[2]{#1 \rightarrow #2}
\newcommand{\fpconds}{F_\text{cond}}
\newcommand{\ppty}[1]{\texttt{(P#1)}}

\newcommand{\plangloballoop}{\mathcal{P}_\texttt{gld}}
\newcommand{\planparallelloop}{\mathcal{P}_\texttt{plw}}
\newcommand{\pgplanparallelloop}{\planparallelloop^\texttt{pg}}
\newcommand{\srplanparallelloop}{\planparallelloop^\texttt{s}}
\newcommand{\rndfile}[3][rnd]{#1\_#3\_#2}
\newcommand{\tree}[1]{\text{tree}\_{#1}}
\newcommand{\uniprot}[1]{uniprot\_#1}
\newcommand{\yago}{\textsc{yago}}
\newcommand{\sizevp}{\mathcal{S}_\text{VP}}

\newcounter{queryCounter}
\newcommand{\samplequery}{~\mathcal{Q}_{\stepcounter{queryCounter}\arabic{queryCounter}}}
\newcounter{uqueryCounter}
\newcommand{\usamplequery}{~\mathcal{Q}_{\stepcounter{uqueryCounter}\arabic{uqueryCounter}}}
\newcommand{\query}[1]{\mathcal{Q}_{#1}}
\newcommand{\querycat}[1]{\mathcal{C}_{#1}}

\newcommand{\checkbox}{\hspace{0.6ex}\times}
\newcommand{\uncheckbox}{}

\newcommand{\setrdd}{\texttt{SetRDD}}
\newcommand{\rdd}{\texttt{RDD}}
\newcommand{\dataset}{\texttt{Dataset}}
\newcommand{\dataframe}{\texttt{DataFrame}}

\newcommand{\anbn}{$a^nb^n$}


\title{Distributed Evaluation of Graph Queries using Recursive Relational Algebra}

\author{\IEEEauthorblockN{Sarah Chlyah\IEEEauthorrefmark{1},
Pierre Genev\`es\IEEEauthorrefmark{2}, Nabil Laya\"ida\IEEEauthorrefmark{3}}
\IEEEauthorblockA{Tyrex team,
Univ. Grenoble Alpes, CNRS, Inria,
Grenoble INP, LIG\\
38000 Grenoble, France\\
Email: \IEEEauthorrefmark{1}sarah.chlyah@inria.fr,
\IEEEauthorrefmark{2}pierre.geneves@inria.fr,
\IEEEauthorrefmark{3}nabil.layaida@inria.fr,
}}

\maketitle

\begin{abstract}
        We present a method and its implementation \distmuir{} for the optimized distributed evaluation of recursive relational algebraic terms. This method provides a systematic parallelisation technique by means of fixpoint splitting, plan generation and selection. The goal is to offer expressivity for high-level queries while providing efficiency and reducing communication costs. Experimental results on both real and synthetic graphs show the effectiveness of the proposed approach compared to existing systems. 
\end{abstract}

\section{Introduction}

With the rise of large-scale graphs in domains like knowledge representation, social networks, and biology \cite{DBLP:journals/cacm/SakrBVIAAAABBDV21}, efficiently extracting information is crucial. This demands scalable methods for distributing data and computations.
Efforts to address these challenges over the past few years have led to various systems such as MapReduce~\cite{dean-osdi04}, Dryad~\cite{dryad2007}, Spark~\cite{ZahariaXWDADMRV16}, Flink~\cite{CarboneKEMHT15} and more specialized graph systems like Google Pregel~\cite{googlepregel2010}, Giraph~\cite{giraph} and Spark GraphX~\cite{GonzalezXDCFS14}.
While these systems can handle large amounts of data and allow users to write a broad range of applications, they still require significant programmer expertise. The system programming paradigms and its underlying configuration tuning must be highly mastered. This includes for example figuring out how to (re)partition data on the cluster, when to broadcast data, in which order to apply operations for reducing data transfers between nodes of the cluster, as well as other platform-specific performance tuning techniques \cite{spark-tuning}.
        
To facilitate large-scale graph querying, it is important to relieve users from having to worry about optimization in the distributed setting, so that they can focus only on formulating domain-specific queries in a declarative manner. A possible approach is to have an intermediate representation of queries (e.g. an algebra) in which high level queries are translated so that they can be optimized automatically. Relational Algebra (RA) is such an intermediate representation that has benefited from decades of research, in particular on algebraic rewriting rules in order to compute efficient query evaluation plans.  
        
A very important feature of graph queries is recursion, which enables to express complex navigation patterns to extract useful information based on connectivity from the graph. For instance, recursion is crucial for supporting queries based on transitive closures. Recursive queries on large-scale graphs can be very costly or even infeasible.
This is due to a large combinatorial of basic computations induced by both the query and the graph topology.
Recursive queries can generate intermediate results that are orders of magnitude larger than the size of the initial graph. For example, a query on a graph of millions of nodes can generate billions of intermediate results. Therefore being able to optimize queries and reduce the size of intermediate results as much as possible becomes crucial.

Several works have addressed the problem of query optimization in the presence of recursion, in particular
with extensions of Relational Algebra \cite{agrawal_alpha:_1988,Aho79,mura-sigmod20}; and with Datalog-based approaches \cite{alice} such as BigDatalog \cite{bigdatalog}.
Recently, $\mu$-RA \cite{mura-sigmod20} proposed logical optimization rules for recursion not supported by earlier approaches.
In particular, these rules include the merging and reversal of recursions that cannot be done neither with Magic sets nor with Demand Transformations that constitute the core of optimizations in Datalog-based systems \cite{mura-sigmod20}. The work in [36] introduces a cost model for [33] that allows for estimating the best logical plan among a set of equivalent \mus{} plans. However, both these works are limited to the centralized setting.

\paragraph*{Contribution} we present \distmuir{}, a new method and its implementation for the optimized distributed evaluation of recursive relational algebra terms. 
The key novelty of our approach lies in the automatic transformation of a global loop into independent local loops, starting from any given term in recursive relational algebra. While systems like Spark and Flink allow for manual implementations of such transformations (e.g., using the mapPartition operator in Spark), they do not perform this transformation automatically.
Specifically, from a high-level query specification (any form compiling into a recursive relational algebra term), our approach enables the automatic optimization of recursions including logical recombinations and transformations into sets of independant local loops that do not require data exchange nor coordination between partitions at each step.
We provide a theoretical guarantee of their correctness. Additionally, we introduce a partitioning strategy that further enhances the efficiency of the evaluation of the independent local loops. The partitioning strategy is formally defined, using the concept of stable columns. One more advantage of this approach is that the criteria for checking whether columns are stable and for enabling and performing the partitioning can be determined statically on an algebraic term.

 Since \distmuir{} implements a generalized relational algebra, it can be of interest for a large number of mainstream RDBMS implementations; and it can also provide the support for distributed evaluation of recursive graph query languages. For example \distmuir{} provides a frontend where the programmer can formulate queries known as UCRPQs \cite{Consens1990,Barcelo2012a,Barcelo2012b,Libkin2016}\footnote{UCRPQs, discussed in more details in Sec.~\ref{sec:architecture}, constitute an important fragment of expressive graph query languages: they correspond to unions of conjunctions of regular path queries. A translation of UCRPQs into the recursive relational algebra is given in \cite{mura-sigmod20}.}).

\section{Preliminaries}\label{sec:preleminaries}
\subsection{\mus{} syntax}

The \mus{} algebra \cite{mura-sigmod20} is an extension of the Codd’s relational algebra with a recursive operator whose aim is to support recursive terms and transform them when seeking efficient evaluation plans. The syntax of \mus{} is recalled from \cite{mura-sigmod20} in Fig.~\ref{fig:syntax}. It is composed of database relation variables and operations (like join and filter) that are applied on relational tables to yield other relational tables. $\mu$ is the fixpoint operator. In $\fixpt{\Psi}$, $X$ is called \emph{the recursive variable} of the fixpoint term.

\begin{figure}[h]
        \begin{small}
        \centering
        \begin{tabular}{cclr}%
                $\varphi$ & $::=$ & ~ & term\\
                &&  $X$ &  relation variable \\
                & $|$ &  $|\cst|$ & constant \\
                & $|$ &  $\varphi_1 \cup \varphi_2$ & union \\
                & $|$ &  $\varphi_1 \NJoin \varphi_2$ & natural join \\
                & $|$ &  $\varphi_1 \triangleright \varphi_2$ & antijoin \\
                & $|$ &  $\filt{\varphi}$ & filtering \\
                & $|$ &  $\rename{a}{b}{\varphi}$ & renaming  \\
                & $|$ &  $\drop{a}{\varphi}$ & anti-projection (column dropping) \\
                & $|$ &  $\fixpt{\Psi}$ & fixpoint term\\
        \end{tabular}\\
        \end{small}
        \caption{Grammar of {\mus} \cite{mura-sigmod20}.}
        \label{fig:syntax}
        \end{figure}

\begin{figure}[h]
        \centering
        \includegraphics[width=7.5cm,keepaspectratio,trim=0cm 1.0cm 0cm 0cm]{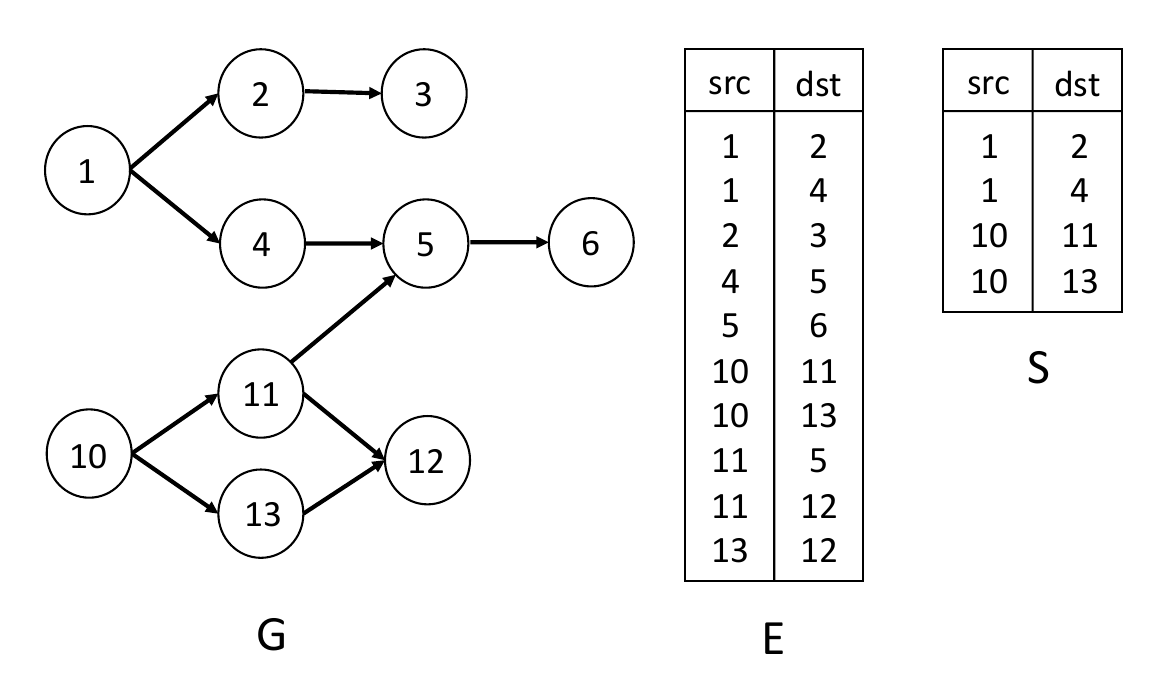} 
        \caption{Graph example.}\label{fig:graphexample}
\end{figure}

Like in RA, the data model in \mus{} consists of relations that are sets of tuples which associate column names to values. For instance, the tuple \{ $src \rightarrow 1$, $dst \rightarrow 2$ \}  is a member of the relation $S$ of Fig.~\ref{fig:graphexample}.

Let us consider a directed and rooted graph $G$, a relation $E$ that represents the edges in $G$, and a relation $S$ of starting edges (a subset of edges in $E$ that start from the graph root nodes), as represented in Fig.~\ref{fig:graphexample}. 

The following examples illustrate how {\mus} algebraic terms can be used to model graph operations, such as navigating through a sequence of edges in a graph:

\begin{ex}\label{ex:simpleexec}
The term $\drop{c}{\rename{dst}{c}{S}\NJoin\rename{src}{c}{E}}$ returns pairs of nodes that are connected by a path of length 2 where the first element of the pair is a graph root node. 
For that purpose, the relation $S$ is joined ($\NJoin$) with the relation $E$ on the common column $c$, after proper renaming ($\rho$) to ensure that $c$ represents both the target node of $S$ and the source node of $E$.  After the join, the column $c$ is discarded by the anti-projection ($\widetilde{\pi}_{c}$) so as to keep only the two columns \texttt{src}, \texttt{dst} in the result relation.
\end{ex}

\begin{ex}\label{ex:fpexec}
Now, the recursive term 
$\fixpt{S \cup \drop{c}{\rename{dst}{c}{X}\NJoin\rename{src}{c}{E}}}$
        computes the pairs of nodes that are connected by a path in $G$ starting from edges in $S$.

The subterm $\varphi = \drop{c}{\rename{dst}{c}{X}\NJoin\rename{src}{c}{E}}$ computes new paths by joining $X$ (the previous paths) and $E$ such that the destinations of $X$ are equal to the sources of $E$.

The fixpoint is computed in 4 steps where $X_i$ denotes the value of the recursive variable at step $i$:
\begin{footnotesize}
\begin{alignat*}{2}
        &X_0&=&\emptyset\\
        &X_1&=&\Big\{\{\mapping{src}{1},\mapping{dst}{2}\}, \{\mapping{src}{1},\mapping{dst}{4}\},\\
        & &&\{\mapping{src}{10},\mapping{dst}{11}\}, \{\mapping{src}{10},\mapping{dst}{13}\}\Big\} \\
        &X_2&=&X_1 \cup \Big\{\{\mapping{src}{1},\mapping{dst}{3}\}, \{\mapping{src}{1},\mapping{dst}{5}\},\\
        & &&\{\mapping{src}{10},\mapping{dst}{5}\},\{\mapping{src}{10},\mapping{dst}{12}\}\Big\}\\
        &X_3&=&X_2 \cup \Big\{\{\mapping{src}{1},\mapping{dst}{6}\},\{\mapping{src}{10},\mapping{dst}{6}\}\Big\}\\
        &X_4&=&X_3 \quad \text{(fixpoint reached)}
\end{alignat*}
\end{footnotesize}
At step 1 it is empty, at step 2 it is a relation of two columns \texttt{src} and \texttt{dst} that contains four rows, and the iteration continues until the fixpoint is reached.
\end{ex}

\subsection {Semantics and properties of the fixpoint}\label{sec:semprop}
The semantics of a \mus{} term is defined by the relation obtained after substituting the free variables in the term (like $E$ and $S$ in example~\ref{ex:fpexec}) by their corresponding database relations. The notions of free and bound variables and substitution are formally defined in \cite{mura-sigmod20}. As a slight abuse of notation, we sometimes use a recursive term $\Psi$ (i.e. a term that contains a recursive variable $X$) as a function $R \to \Psi(R)$ that takes a relation $R$ and returns the relation obtained by replacing $X$ in the term $\Psi$ by the relation $R$. In the above example 

\begin{footnotesize}
$\varphi(S) = \drop{c}{\rename{dst}{c}{S}\NJoin\rename{src}{c}{E}} = \Big\{\{\mapping{src}{1},\mapping{dst}{3}\}, \{\mapping{src}{1},\mapping{dst}{5}\},\{\mapping{src}{10},\mapping{dst}{5}\},\{\mapping{src}{10},\mapping{dst}{12}\} \Big\} $.
\end{footnotesize}

Under this notation, $\fixpt\Psi$ is defined as the fixpoint $F$ of the function $\Psi$, so $\Psi(F)=F$.

\begin{defirep}{Let us consider the following conditions, denoted $\fpconds$, for a fixpoint term $\fixpt{\Psi}$:}
\begin{packeditemize}
        \item \emph{positive}: for all subterms $\varphi_1\AJoin \varphi_2$ of $\Psi$, $\varphi_2$ is constant in $X$ (i.e. $X$ does not appear in $\varphi_2$);
        \item \emph{linear}: for all subterms of $\Psi$ of the form $\varphi_1\NJoin \varphi_2$ or $\varphi_1\AJoin \varphi_2$, either $\varphi_1$ or $\varphi_2$ is constant in $X$;
        \item \emph{non mutually recursive}: when there exists a subterm $\fixpt[Y]{\psi}$ in $\Psi$, then any occurence of $X$ in this subterm should be inside a term of the form $\fixpt\gamma$.
\end{packeditemize}
\end{defirep}

These conditions guarantee the following properties (see \cite{mura-sigmod20}): 
\begin{small}
\begin{proprep}
        \label{prop:fixpt}
        If $\fixpt{\Psi}$ satisfies $\fpconds$ then \[\Psi(S)= \Psi(\emptyset) \cup \bigcup_{x\in S} \Psi(\{x\}) \]
        and thus $\Psi$ has a fixpoint with
        $\fixpt{\Psi}=\Psi^{\infty}(\emptyset)$. 
\end{proprep}
\end{small}
        For instance, $\fixpt{R \AJoin X}$ is not positive, $\fixpt{X \Join X}$ is not linear, and $\fixpt{\fixpt[Y]{\varphi(X)}}$ is mutually recursive. Whereas $\fixpt{R \cup X \Join \fixpt[Y]{\varphi(Y)}}$ satisfies $\fpconds$.

\begin{proprep}
\label{prop:decfixpt}
Every fixpoint term $\fixpt\Psi$ that satisfies $\fpconds$ can be written like the following: $\fixpt{R \cup \varphi}$ where $R$ is constant in $X$ and $\varphi(\emptyset)=\emptyset$. $R$ is called \textbf{the constant part} of the fixpoint and $\varphi$ \textbf{the variable part}.
\end{proprep}
In Example~\ref{ex:fpexec}, $S$ is the constant part and $\drop{c}{\rename{dst}{c}{X}\NJoin\rename{src}{c}{E}}$ is the variable part.
In the rest of the paper, we only consider fixpoint terms satisfiying the conditions $\fpconds$ in their decomposed form  $\fixpt{R \cup \varphi}$ since their existence is guaranteed thanks to proposition~\ref{prop:fixpt}.

The evaluation of recursive terms has been studied in the context of Datalog \cite{alice} and with transitive closure evaluation \cite{yannis86} with the semi-naive (or differential) method. In this approach, a fixpoint term is typically evaluated with the algorithm~\ref{alg:eval}.

\lstset{
        basicstyle=\footnotesize\ttfamily, 
        numberstyle=\small,          
        numbersep=5pt,              
        tabsize=2,                  
        extendedchars=true,
        breaklines=true,            
        keywordstyle=\color{red},
        stringstyle=\color{white}\ttfamily, 
        showspaces=false,
        showtabs=false,
        xleftmargin=17pt,
        framexleftmargin=17pt,
        framexrightmargin=5pt,
        framexbottommargin=4pt,
        showstringspaces=false
}

\begin{small}
\begin{algorithm}[H]
\begin{lstlisting}
    X = R 
    new = R 
    while new $\neq$ $\emptyset$:
        new = $\varphi$(new)$\setminus$ X
        X = X $\cup$ new
    return X
\end{lstlisting}
\caption{}\label{alg:eval}
\end{algorithm}	
\end{small}

The final result is obtained by evaluating $\varphi$ repeatedly starting from $X=R$ until no more results can be produced (the fixpoint has been reached). In this algorithm, $\varphi$ is applied on the new results only (obtained by making a set difference between the current result and the previous one) instead of the entire result set.  Notice that this is possible thanks to the property of $\varphi$ stated in proposition~\ref{prop:fixpt}, which implies that $\varphi(X_i) \cup \varphi(X_{i+1}) = \varphi(X_i) \cup \varphi(X_{i+1} \setminus X_i)$.

The work found in \cite{mura-sigmod20} that introduced \mus{}  considers only the centralized setting. In this paper, we address the evaluation of recursive terms in a distributed manner.

\section{Fixpoint parallelization and distribution}\label{sec:distribution}
In this section, we extend \mus{} to enable efficient distributed evaluation.
We first introduce the principles of how we achieve parallelization of fixpoint terms and then describe the implementation on a specific distributed framework: Spark \cite{ZahariaXWDADMRV16}.

\subsection{Fixpoint parallelization principle}\label{sec:parallelization}
We propose a parallelization technique based on the following proposition:

\begin{proprep}{Fixpoint Splitting.}
        Under the aforementioned conditions $\fpconds$, we have:
         \label{prop:distribution}$$\fixpt{R_1 \cup R_2 \cup \varphi} = \fixpt{R_1 \cup \varphi} \cup \fixpt{R_2 \cup \varphi}$$
\end{proprep}
 This proposition means that a fixpoint whose constant part is a union of two datasets can be obtained by making the union of two fixpoints, each with one of these datasets as a constant part.
We can leverage this property for parallelizing the fixpoint computation by dividing the constant part $R$ into $n$ portions $R_i$ and computing $n$ smaller fixpoints $\fixpt{R_i\cup\varphi}$ in parallel. The results of those fixpoints are then combined using the union operator $\cup$ that eliminate duplicates.

In Example \ref{ex:fpexec}, if we split the start edges $S$ into $S_1$ = $\{(1,2), (10,11)\}$ and $S_2=\{(1,4), (10,13)\}$, we can compute the following sets of paths in parallel $\{(1,3),(10,5), (10,6), (10,12)\}$ and $\{(1,5), (1,6), (10,12)\}$.

\subsection{Data partitioning}\label{sec:fpdistrepartition} 

Splitting the initial constant part $R$ into several portions $R_i$ may produce duplicates accross the computed subresults that are then eliminated using the $\cup$ operator. 
However, for some cases, an appropriate partitioning of data produces parallel subresults which are disjoint. 
We present a criterion that can be automatically verified and a partitioning strategy that ensures the final duplicate elimination step can be omitted when applied to expressions meeting this criterion.
We first present the intuition behind this idea, followed by the proof.

We look for a column $col$ (or a set of columns) in $X$ that is left unchanged by $\varphi$. In other words, a tuple in $R$ having a value $v$ at column $col$ will only generate tuples having the same value at this column throughout the iterations of the fixpoint. So if we put all tuples in $R$ having $v$ at column $col$ in one partition, no other partition will generate a tuple with this value at that column.  
For this, we first compute the set of \emph{stable} columns which are not altered during the fixpoint iteration. For instance, in the fixpoint expression of example~\ref{ex:fpexec}, 'src' is a \emph{stable} column. This means that tuples in the fixpoint having 'src' = 1 can only be produced from tuples in S having 'src' = 1. This implies that filtering tuples having 'src' = 1 before or after the fixpoint computation lead to the same results.
However, this is not true for the column 'dst' which is not \emph{stable}. 
To summarize, when the constant part of the fixpoint is partitioned by the stable column (or columns)\footnote{$R$ is partitioned by a column $c$ when $c$ does not take the same value on two different partitions $R_i$ and $R_j$.} prior to the fixpoint execution, we know for certain that there will be no duplicate across the parallel subresults. 
For instance with the partitioning $S_1=\{(1,2), (1,4)\}$ and $S_2=\{(10,11), (10,13)\}$, which corresponds to a partitioning of $S$ by the column 'src', we obtain the following sets of paths $\{(1,3), (1,5), (1,6)\}$ and $\{(10,5), (10,6), (10,12)\}$, thus avoiding the duplicate $(10,12)$ obtained with the previous partitioning. 
\begin{inlineproof}
        Let $c$ be a stable column of $\fixpt{R \cup \varphi}$, which means that $\forall e \in \fixpt{R \cup \varphi} \;\;  \exists r \in R \; \; e(c) = r(c)$ \cite{mura-sigmod20}. In \mus{}, an element $r$ in $R$ is a mapping (tuple), which means that it is a function that takes a column name and returns the value that $r$ has at that column.

        Let us consider a partitioning $R_1$, ..., $R_n$ of $R$ by the column $c$ which verifies the following 
        $$\forall i \neq j \in \{1..n\} \; \forall a \in R_i \; \forall b \in R_j \;\; a(c) \neq b(c) $$
        This statement means that there are no two elements of $R$ at different partitions that share the same value at column $c$. We next show that this statement is also true for the fixpoint term.
        
        Let $i \neq j \in \{1 ..n\}$ and let $x \in \fixpt{R_i \cup \varphi}$ and $y \in \fixpt{R_j \cup \varphi}$. Since $c$ is stable, we have $\exists a \in R_i \;\; x(c) = a(c)$ and $\exists b \in R_j \;\; y(c) = b(c)$. So $x(c) \neq y(c)$. 
                
        In conclusion, the sets $\fixpt{R_i \cup \varphi} $ where $i \in \{1 .. n\}$ are disjoint.
\end{inlineproof}

        \paragraph*{Applicability of data partitioning} 
        The partitioning technique relies on the presence of a stable column, which is a necessary condition for its application. 
        However, for any recursive subterm in a UCRPQ, a stable column always exists, ensuring that the partitioning technique can always be applied.
        This guarantee arises because, for each recursion in a UCRPQ query, the translation and the rewriting automatically generate two algebraic plans: one that evaluates recursion left to right, and one that evaluates recursion right to left.
        Since a column that is not stable in one plan is necessarily stable in the other, this ensures that a stable column is always available, making partitioning feasible for any UCRPQ.
        %
    %
    In the given example for instance, recursion is performed from left to right which makes the column 'src' stable. In the equivalent plan that performs it from right to left, which is also generated, the column 'trg' is stable.
    %
        Furthermore, the technique can apply to non-regular queries (beyond UCRPQ).
        For instance, in a fixpoint defining a non-regular pattern such as $a^nb^n$, the relations $a$ and $b$ may contain additional columns beyond 'src' and 'trg', making them stable. In such cases, the partitioning technique remains applicable. However, if such a non-regular fixpoint does not contain any additional columns (beyond 'src' and 'trg'), then the proposed technique cannot apply, due to the lack of stable columns.

\subsection{Fixpoint distributed evaluation on Spark}\label{sec:fpdistrib}

We now describe how fixpoint terms are evaluated in a distributed manner on the Spark~\cite{ZahariaXWDADMRV16} plaform.

We consider two ways of distributing the fixpoint computation in a Spark cluster: (1) $\plangloballoop$ which 
 uses a global loop on the Spark \emph{driver}\footnote{The \emph{driver} is the process that creates tasks and send them to be executed in parallel by \emph{worker} nodes.}. (2) $\planparallelloop$ which uses parallel local loops on the Spark workers.

\subsubsection{\underline{G}lobal \underline{L}oop on the \underline{D}river ($\plangloballoop$)}
\begin{figure}[h]
        \centering
        \includegraphics[keepaspectratio,width=0.38\textwidth]{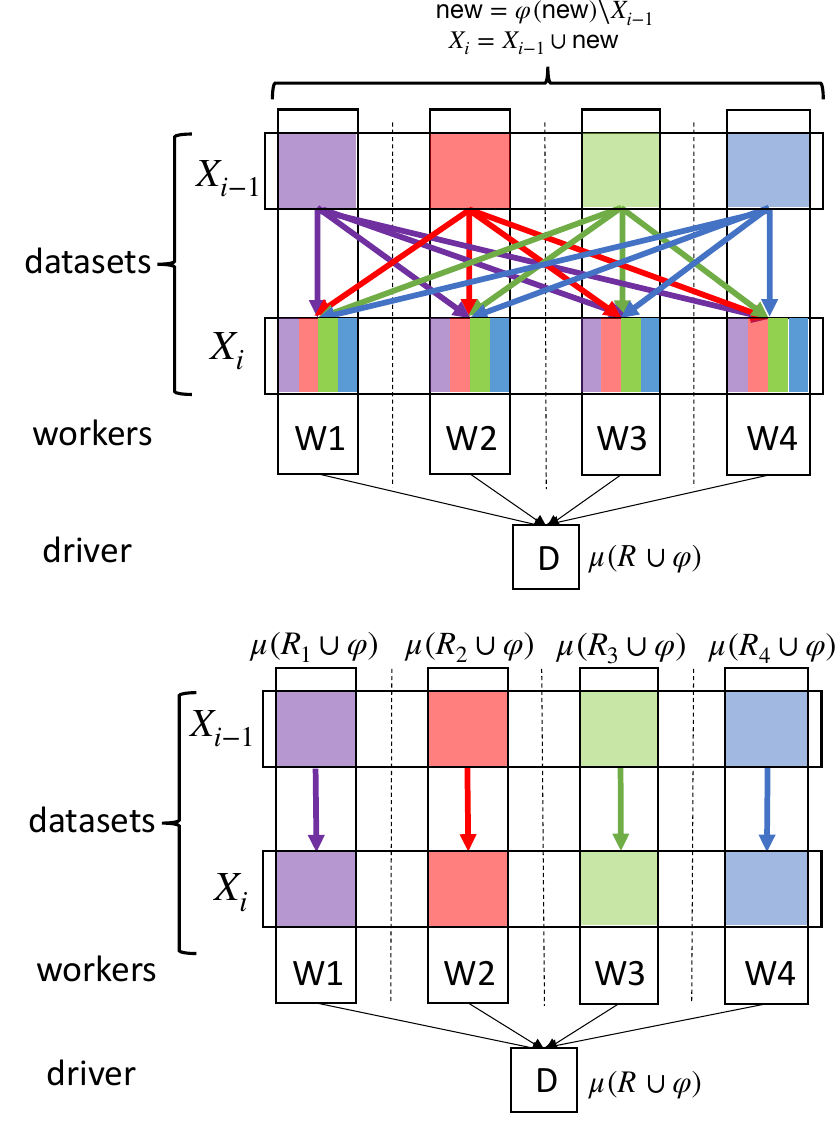}
        \caption{Distributed execution of $\plangloballoop$ (top) \& $\planparallelloop$ (bottom).}\label{fig:muplans}
        \end{figure}

$\plangloballoop$ corresponds to the natural way a Spark programmer would implement the fixpoint operation: it distributes the computations performed at each iteration of Algorithm~\ref{alg:eval}. This execution is illustrated in Fig.~\ref{fig:muplans} (top part). 

Colored arrows show data transfers that occur at each iteration of the fixpoint. The driver performs the loop and, at each iteration, instructions at lines 4 and 5 are executed as \dataset{} operations \cite{sparksql2015} that are distributed among the workers. We call this execution plan $\plangloballoop$. On Spark, $\cup$ is executed as \dataset{} union followed by a \texttt{distinct()} operation. This means that in $\plangloballoop$, at least one data transfer (shuffle) per iteration is made to perform the union. 

\subsubsection{\underline{P}arallel \underline{L}ocal loops on the \underline{W}orkers ($\planparallelloop$)}
$\planparallelloop$ corresponds to the parallelization strategy presented in Sec.~\ref{sec:parallelization}. The driver splits the constant part $R$ among the workers, then each worker $i$ executes the fixpoint $\fixpt{R_i\cup\varphi}$ locally starting from its own constant part $R_i$. We call this execution plan $\planparallelloop$. Execution is illustrated in the bottom part side of Fig.~\ref{fig:muplans}.  
As opposed to $\plangloballoop$, $\planparallelloop$ performs only one data shuffle at the end to make the union ($\cup$) between the local fixpoints.
This final shuffle can also be avoided by appropriately repartitioning input data among workers as explained in Sec.~\ref{sec:fpdistrepartition}.

\subsection{Physical plan generation and selection}\label{sec:physicalplanning}

We present the different physical plans automatically generated by \distmuir{}, and explain how they are selected. \distmuir{} generates a physical plan for $\plangloballoop$, which is used only as a baseline in performance comparisons.  
\begin{figure}
\includegraphics[keepaspectratio,width=0.38\textwidth]{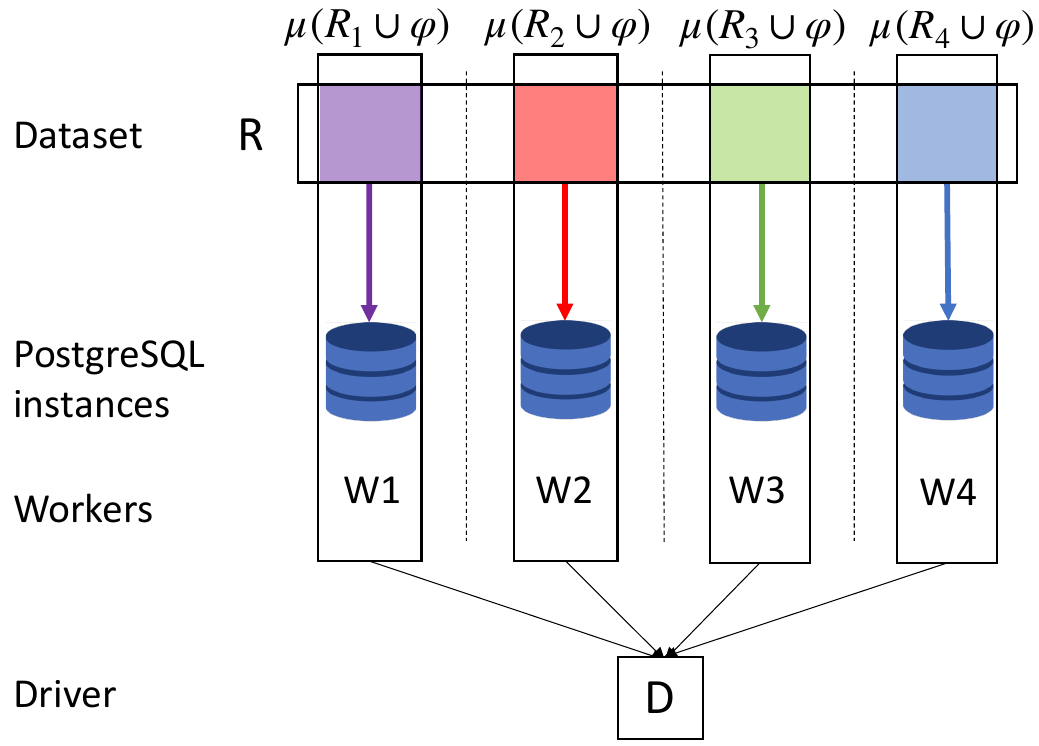}
\\
\includegraphics[keepaspectratio,width=0.37\textwidth]{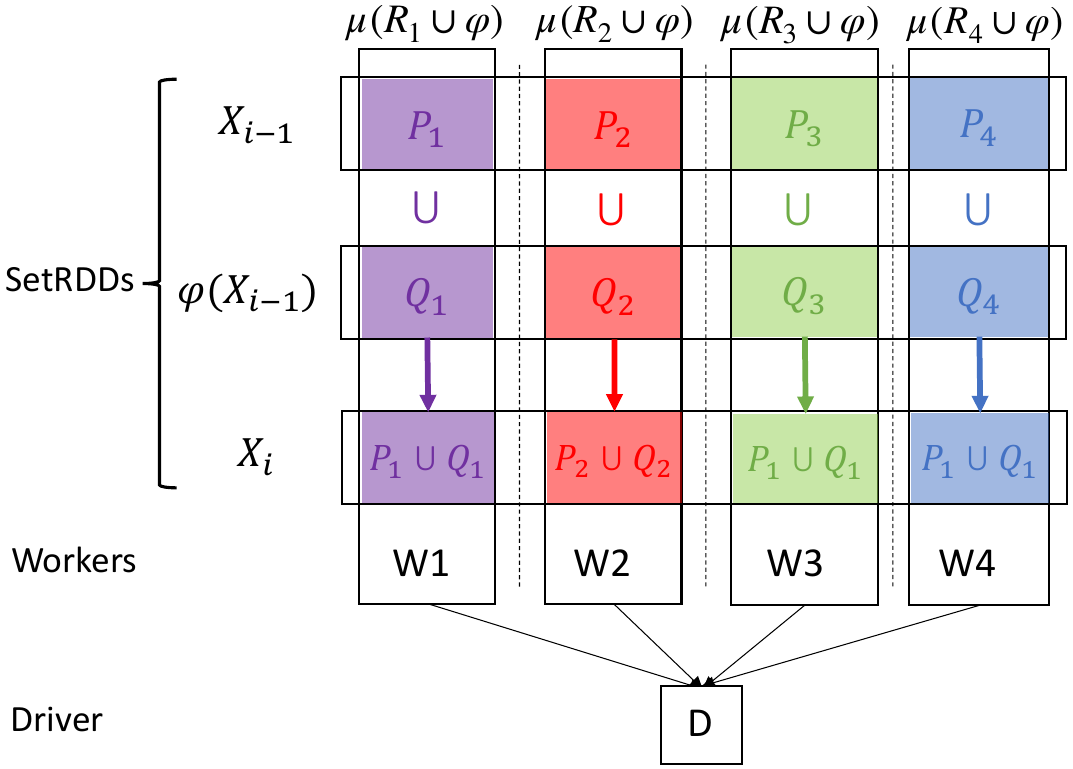}
\caption{Distributed execution of $\planparallelloop$ plans.}\label{fig:pg}
\end{figure}
We propose two alternative physical plans which are variants of $\planparallelloop$:  
\begin{packeditemize}
        \item $\pgplanparallelloop$: The top of Fig.~\ref{fig:pg} illustrates the execution of this physical plan. The local fixpoints are executed on PostgreSQL.
        The fixpoint operator is performed as a Spark \texttt{mapPartition()} operation where each worker performs a portion of the fixpoint computation on PostgreSQL. A PostgreSQL instance runs on each worker. The part of data assigned to each worker is represented as a view in the PostgreSQL instance running on this worker. The \mus{} expression (that computes the fixpoint) is translated to a PostgreSQL query that is executed using this view as the constant part of the fixpoint.  The local PostgreSQL plans are selected for the operators in the fixpoint expression. 
        Each PostgreSQL executor returns its results as an iterator which is then processed by Spark.

        \item $\srplanparallelloop$: The bottom of Fig.~\ref{fig:pg} illustrates the execution of this physical plan. The fixpoint computation is implemented using a loop in the driver that uses Spark operations to compute the recursive part of the fixpoint. These Spark operations are written in such a way that each worker performs its own fixpoint independently (i.e. without data exchanged between workers). Joins are executed as broadcast joins: all relations in the variable part of the fixpoint (apart from the recursive relation) are broadcasted. Antiprojections are executed without the need of applying the \texttt{distinct()} operation. To perform the union (or set-difference), a special union (set-difference) operation is used that computes the union (set-difference) partition-wise. These special union and set-difference operations are implemented as part of the \setrdd{} API. \setrdd{}  \cite{bigdatalog} is a special \rdd{}\footnote{\rdd{} is an abstraction that Spark provides to represent a distributed collection of data. An \rdd{} is split among partitions which are assigned to workers.} where each partition is a set. This \setrdd{} is used to store the value of the recursive variable X at each iteration. This means that each partition of X holds the intermediate results of the local fixpoint performed by the worker to which this partition has been assigned. 
\end{packeditemize}

As a consequence, for each of the non-recursive \mus{} operators, there are two kinds of physical plans: local plans implemented using PostgreSQL and distributed plans implemented using the Spark \dataset{} API. \dataset{}s are used to represent relational data in Spark. The optimization of these expressions is then delegated to Spark's Catalyst internal optimizer \cite{sparksql2015} before execution.
Some operators have more than one distributed execution plan. For instance, for the join operator, we choose which argument (if any) to broadcast in order to guide Spark on whether to use broadcast join or another type of join.

\begin{figure}
        \includegraphics[trim={0 0cm 0 0},clip]{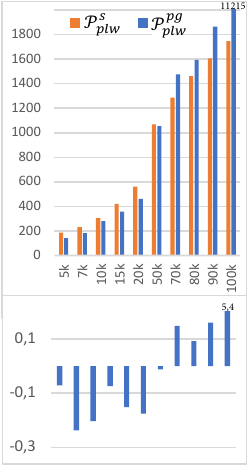}                
        \includegraphics[trim={0 0cm 0 0},clip]{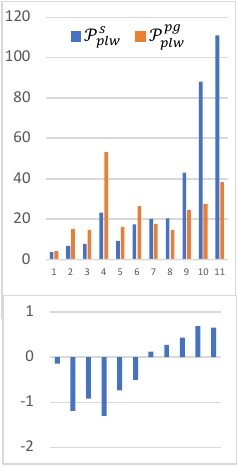}                
\caption{Comparison between $\pgplanparallelloop$ and $\srplanparallelloop$.}\label{fig:newcomparisonbetweenplans}
\end{figure}

\paragraph*{Selecting between $\planparallelloop$ variants}     
In $\srplanparallelloop$, datasets in the variable part of the fixpoint are broadcasted to the workers, whereas in $\pgplanparallelloop$, they are are stored as tables in Postgres that are queried by the parallel tasks. 
Experimental observations indicate that the relative performance between $\pgplanparallelloop$ and $\srplanparallelloop$ is impacted by the size of intermediate data during fixpoint computations (i.e., the subterm $\phi(X)$ in a fixpoint $\fixpt{c \cup  \phi(X)}$. This size is determined by (1) the size of the fixpoint constant part and (2) the size of constant subterms in $\phi(X)$.
To quantify the impact of the two factors, we first compare the $\pgplanparallelloop$ and $\srplanparallelloop$ plans on a transitive closure computation with varying constant part sizes. Results are shown in Fig.~\ref{fig:newcomparisonbetweenplans} (left) for a randomly generated Erdős-Rényi graph with 10M edges, where the size of the constant part, which ranges from 5k to 100k records, is indicated on the $x$-axis, and evaluation time (in seconds) on the $y$-axis. For the second factor, we make both plans navigate the Yago graph using a Kleene star over various expressions that start from a single node. Each expression under the Kleene star links a number of node pairs ('src','trg') which varies from 10,000 to 70 million records. Results are shown in Fig.~\ref{fig:newcomparisonbetweenplans} (right), where the $x$-axis indicates query indexes ranked in increasing order of $\phi(X)$ size (corresponding queries can be found in \cite{distmura-page}), and the $y$-axis shows evaluation time (in seconds).
Associated with the top charts of Fig.~\ref{fig:newcomparisonbetweenplans} are the bottom charts that show the respective speedup (on the $y$-axis) of $\pgplanparallelloop$ compared to $\srplanparallelloop$.
Experimental results suggest that when the intermediate data is of relatively small to moderate size, $\pgplanparallelloop$ is impacted by the overhead of data marshalling and transfer between Spark and PostgreSQL. When the size of intermediate data becomes larger, $\pgplanparallelloop$ starts benefitting from PostgreSQL's highly optimized implementation including efficient index usage and optimized complex joins. 
Based on these observations, we adopt a simple heuristic for choosing between the two implementations: when the size of the datasets in the variable part of the fixpoint exceeds the memory available for a task\footnote{A Spark task is
unit of computation executed on the worker on a single partition. Tasks are executed in parallel on partitioned data.}, we select $\pgplanparallelloop$ and $\srplanparallelloop$ otherwise. 

\section{\distmuir{} architecture}\label{sec:architecture}

\begin{figure}[h]
        \centering
        \includegraphics[width=9cm,keepaspectratio]{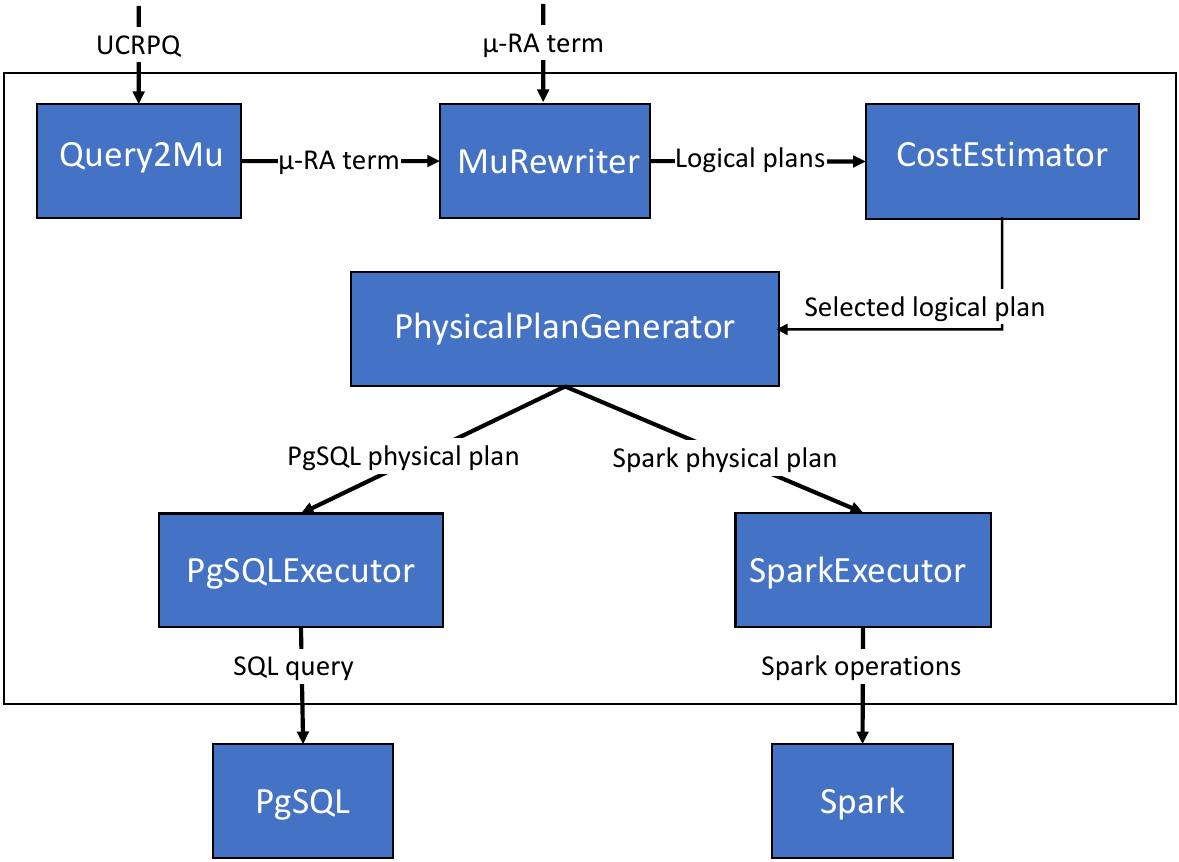}
        \caption{Architecture of the \distmuir{} system.}\label{fig:architecture}
\end{figure}

The \distmuir{} system takes a query as input parameter, translates it into {\mus}, optimizes it, and then performs the evaluation in a distributed fashion on top of Spark. Specifically, the \distmuir{} system is composed of several components, as illustrated in Fig.~\ref{fig:architecture}.

The \texttt{Query2Mu} component translates recursive graph queries written in Union of Conjunctive Regular Path Queries (UCRPQ) into \mus{} terms. The UCRPQ syntax is given in~\cite{mura-sigmod20} and we give an example below.  \distmuir{} supports more general \mus{} terms that are not expressible as UCRPQs\footnote{See the practical experiments section for some examples such as the ``same generation'' query.}, as long as they satisfy the triple condition $\fpconds$ mentioned in Section~\ref{sec:preleminaries}. 

From a given input \mus{} term, the \texttt{MuRewriter} explores the space of semantically equivalent logical plans by applying a number of rewrite rules. In addition to the rewrite rules already known in classical relational algebra, \texttt{MuRewriter} applies a set of rules specific to the fixpoint operator. These rules and the conditions under which they are applicable are formally defined in \cite{mura-sigmod20}. 

The evaluation costs of these terms are estimated by the \texttt{CostEstimator} component proposed in \cite{muideen-cikm2020}. This component is an implementation of a classical Selinger style cost estimator~\cite{SelingerACLP79} based on cardinality estimation for subterms. It uses cardinalities estimated by PostgreSQL for base relations; standard cost estimation techniques for non-recursive constructs; and the logarithm-based estimation technique proposed in~\cite{lawal:tel-03322720} for fixpoints. Based on these estimations, a best logical recursive plan is selected.

From a given recursive logical plan, the \texttt{PhysicalPlanGenerator} generates a physical plan for distributed execution (see Section~\ref{sec:distribution}).
Two distributed execution setups are used.
In the first (using \texttt{PgSQLExecutor}), each Spark worker runs a PostgreSQL instance to perform a part of the evaluation locally. The second setup (using \texttt{SparkExecutor}) relies only on Spark. In all cases, the query evaluation is performed on top of Spark.

\paragraph{Example}

We first describe the transformations that a query (UCRPQ or $\mu$-RA term) undergoes before being considered for distributed evaluation when given as input parameter to the \texttt{PhysicalPlanGenerator} (described in Section~\ref{sec:distribution}).

Consider for instance the following UCRPQ composed of a conjunction of two Regular Path Queries (RPQs):
\begin{flalign*}
\texttt{?a,?b,?c}  \leftarrow  &\texttt{?a wasBornIn/IsLocatedIn+ Japan,}\\ 
& \texttt{?b isConnectedTo+ ?c}
\end{flalign*}
The first RPQ computes the people \texttt{?a} that are born in a place that is located directly or indirectly in \texttt{Japan}.   
The query is first translated into \mus{} by \texttt{Query2Mu} so that \texttt{MuRewriter} can generate semantically equivalent plans. We describe below the rewrite rules specific to fixpoint terms leveraged from \cite{mura-sigmod20} that can apply in \texttt{MuRewriter}, and we give the intuition of their effect on performance: 
\begin{packeditemize}
        \item \emph{Pushing filters into fixpoints}: with this rule, the query \texttt{?x}~\texttt{isLocatedIn}+~\texttt{Japan} is evaluated as a fixpoint starting from \texttt{?x} such as \texttt{?x isLocatedIn Japan}, which avoids the computation of the whole \texttt{isLocatedIn+} relation followed by the filter \texttt{Japan}. 
        
        \item \emph{Pushing joins into fixpoints}: let us consider the query \texttt{?x isMarriedTo/knows+ ?y}. Instead of computing the relation \texttt{knows+} and joining it with \texttt{isMarriedTo}, this rule rewrites the fixpoint such that it starts from \texttt{?x} and \texttt{?y} that verify \texttt{?x isMarriedTo/knows ?y}. The application of this rule is beneficial in this case because the size of the \texttt{isMarriedTo/knows} relation is usually smaller than the size of the knows relation.

        \item \emph{Merging fixpoints}: when evaluating \texttt{?x isLocatedIn+/dealsWith+ ?y}, instead of computing both fixpoints separately and then joining them, this rule generates a single fixpoint that starts with \texttt{isLocated/dealsWith} then recursively appends either \texttt{isLocatedIn} to the left or \texttt{dealsWith} to the right. 
        
        \item \emph{Pushing antiprojections into fixpoints}: this rule removes unused columns during the fixpoint computations. For instance, the query \texttt{?y $\shortleftarrow$ ?x isLocatedIn+ ?y} (this query asks for \texttt{?y} only) is evaluated by starting from the destinations \texttt{?y} of the \texttt{isLocatedIn} relation and recursively retrieving new destinations, thereby avoiding the need to store pairs \texttt{(?x,?y)} only to discard \texttt{?x} later.

        \item \emph{Reversing a fixpoint}: the fixpoint corresponding to the relation $a+$ can either be computed from left to right by starting from $a$ and by recursively appending $a$ to the right of the previously found results, or from right to left by starting from $a$ and appending $a$ to the left. Reversing a fixpoint consists in rewriting from the first form to the other or vice versa. This rule is necessary to account for all possible filters and joins that can be pushed in a fixpoint. For instance, a filter that is located at the left side of $a+$ can only be pushed if the fixpoint is evaluated from left to right.
\end{packeditemize}

After these transformations, the best (estimated) recursive logical plan selected by \texttt{CostEstimator} is given as input parameter to \texttt{PhysicalPlanGenerator} that is in charge of generating the best physical plan for distributed execution.

\section {Experiments}\label{sec:experiments}

We evaluate the performance of a prototype implementation of the \distmuir{} system on top of the Spark platform \cite{ZahariaXWDADMRV16}. 
We extensively compared its performance against other state-of-the-art systems on various datasets and queries. We report below on these experiments.

\subsection{Experimental setup} \label{sec:xpsettings}
Experiments were conducted on a Spark cluster with four machines, each running a worker and one also hosting the driver.  Each machine has 40~GB RAM, 2 Intel Xeon E5-2630 v4 CPUs (2.20 GHz, 20 cores each) and 66 TB of 7200 RPM HDDs, running Spark 2.4.5 and Hadoop 2.8.4 in Debian-based Docker containers.

\subsection{Datasets} 

We use real and synthetic datasets of different sizes and topological properties (as detailed in \cite{distmura-page}). We consider the following real graphs:
\begin{packeditemize}
    \item{Yago}\footnote{\label{footnote:yago}We use a cleaned version of the real world dataset Yago 2s, that we have preprocessed in order to remove duplicate RDF \cite{rdf} triples (of the form <source, label, target>) and keep only triples with existing and valid identifiers. After preprocessing, we obtain a table of Yago facts with 83 predicates and 62,643,951 rows (graph edges).}: A knowledge graph extracted from Wikipidia  \cite{yago}. 
    \item datasets from the Colorado index of complex networks \cite{icon} and from the Snap network dataset collection \cite{snap}.
\end{packeditemize}

In addition, we consider the following synthetic graphs:
\begin{packeditemize}
    \item \uniprot{n}: a benchmark graph of $n$ nodes generated using the gMark benchmark tool \cite{bagan-tkde2017}. It models the Uniprot database of proteins \cite{uniprotdb}.
    \item \rndfile{p}{n}: random graphs generated with the Erdos Renyi algorithm, where $n$ is the number of nodes in the graph and $p$ the probability that two nodes are connected.
    \item \tree{n}: a random tree of $n$ nodes generated recursively as follows: \tree{1} is a tree of 1 node, and then $\tree{i+1}$ is a tree of $i+1$ nodes where the $i^\text{th}+1$ node is connected as a child of a randomly selected node in $\tree{i}$.
\end{packeditemize}

\subsection{Systems}\label{sec:xpsystems}
 We  compare \distmuir{} with the following systems:
\begin{packeditemize}
    \item BigDatalog~\cite{bigdatalog} available at \cite{bdl-github}: a large-scale distributed Datalog engine built on top of Spark.
    \item GraphX~\cite{GonzalezXDCFS14}: a Spark library for graph computations. It exposes the Pregel API for recursive computations. 
    In order to compare our system with GraphX we need to convert UCRPQs to GraphX programs\footnote{In the GraphX framework, a recursive computation is composed of ``supersteps'' where, in each superstep, graph nodes send messages to their neighbor nodes, then a merge function aggregates messages per recipient and each recipient receives its aggregated messages in order to process them. A computation is stopped when no new message is sent.}.
    Specifically, we compute a regular graph query by making each node send a message to its neighbors in such a way that the query pattern is traversed recursively from left to right. This means that for a query that starts by selection \texttt{(?x $\leftarrow$ A pattern ?x)}, only the node \texttt{A} sends a message at the start of the computation. 
\end{packeditemize}
    \subsection{Queries}\label{xp:queries}

Queries may involve different forms of recursion. To ensure diverse coverage, we classify them into seven categories. Classes $\querycat{1}-\querycat{6}$ cover UCRPQ queries on knowledge graphs (e.g., Uniprot, Yago), while $\querycat{7}$ includes more general queries beyond UCRPQs. The classification is as follows:

\begin{packeditemize}

    \item $\querycat{1}$ corresponds to queries containing a single transitive closure (TC), e.g. $\texttt{?x}, \texttt{?y} \leftarrow \texttt{?x a+ ?y}$ 
    \item $\querycat{2}$: queries with a filter to the right of a TC, e.g. $\texttt{?x} \leftarrow \texttt{?x a+ C}$ 
    \item $\querycat{3}$: queries with a filter to the left a TC, e.g. $\texttt{?x} \leftarrow \texttt{C a+ ?x}$
    \item $\querycat{4}$: queries which contain a concatenation of a non recursive term to the right of a TC, e.g. $\texttt{?x}, \texttt{?y} \leftarrow \texttt{?x a+/b ?y}$
    \item $\querycat{5}$: queries which contain a concatenation of a non recursive term to the left of a TC, e.g. $\texttt{?x}, \texttt{?y} \leftarrow \texttt{?x b/a+ ?y}$
    \item $\querycat{6}$: queries which contain a concatenation of TCs, e.g. $\texttt{?x},~\texttt{?y}~\leftarrow~\texttt{?x~a+/b+~?y}$
    \item $\querycat{7}$: queries with non regular recursion, e.g. \anbn.
\end{packeditemize}

Each class requires specific optimizations. For instance, the optimization of queries of classes $\querycat{2}$ and $\querycat{3}$ requires pushing filters in fixpoint terms (in two different directions). Queries of classes $\querycat{4}$ and $\querycat{5}$ require an optimization that pushes joins in fixpoint terms. $\querycat{2}$ and $\querycat{4}$ require reversing fixpoint terms before applying other optimizations (rewritings). Queries of $\querycat{6}$ can be optimized by merging fixpoints or by pushing joins in  fixpoint terms.

A query may belong to one or more classes. When a query belongs to multiple classes, it requires the optimization techniques from all corresponding classes, along with a method to combine them. Therefore, the more classes a query belongs to, the more challenging its optimization.
For example, the query $\texttt{?x} \leftarrow \texttt{C a/b+ ?x}$ belongs to $\querycat{3}$ because there is a filter to the left of the transitive closure \texttt{b+} and also belongs to $\querycat{5}$ because there is a concatenation to the left of \texttt{b+}.

To cover a variety of queries in the experiments (see Figures~\ref{fig:yagoQueries} and \ref{fig:uniprotQueries2}), there is, for each class $\querycat{i}$, at least one query that belongs to $\querycat{i}$ alone. In addition, we also consider queries that belong to $\querycat{i}$ and to a combination of other classes. 
This allows to test how the different combinations of optimizations are supported by the tested systems.

\paragraph{Yago queries} Fig.~\ref{fig:yagoQueries} lists UCRPQs evaluated on the Yago dataset along with their classes.
Queries $\query{3}$ and $\query{4}$ are taken from \cite{abul-edbt17}, $\query{5}$ from \cite{yakovets2015waveguide}, and $\query{6},\query{7}$ from \cite{gubichev-grades13}. We have added queries $\query{8}-\query{25}$ that include larger transitive closures. 
\begin{figure}[h]
    \begin{center}
    \begin{tiny}
        \setttsize{\tiny}
    $\begin{array}{|@{\hskip 1pt}l@{\hskip 1pt}|l@{\hskip 1pt}l@{\hskip 1pt}l|@{\hskip 1pt}l@{\hskip 1pt}l@{\hskip 1pt}l@{\hskip 1pt}l@{\hskip 1pt}l@{\hskip 1pt}l@{\hskip 1pt}|} \hline 
        \query{id}&Query&&&\querycat{1}&\querycat{2}&\querycat{3}&\querycat{4}&\querycat{5}&\querycat{6}\\ \hline \rule{0pt}{3ex}  
    \samplequery& \texttt{?x,?y}  &\leftarrow&   \texttt{?x,?y <- ?x hasChild+ ?y}  & \checkbox{}&\uncheckbox{}&\uncheckbox{}&\uncheckbox{}&\uncheckbox{}&\uncheckbox{}  \\
    \samplequery& \texttt{?x,?y}   &\leftarrow&    \texttt{?x,?y <- ?x isConnectedTo+ ?y} &  \checkbox{}&\uncheckbox{}&\uncheckbox{}&\uncheckbox{}&\uncheckbox{}&\uncheckbox{}\\
    \samplequery& \texttt{?x}   &\leftarrow&   \texttt{?x isMarriedTo/livesIn/IsL+/dw+ Argentina} &  \uncheckbox{}&\checkbox{}&\uncheckbox{}&\uncheckbox{}&\checkbox{}&\checkbox{}\\
    \samplequery& \texttt{?x}   &\leftarrow&   \texttt{?x livesIn/IsL+/dw+ United\_States} &  \uncheckbox{}&\checkbox{}&\uncheckbox{}&\uncheckbox{}&\checkbox{}&\checkbox{}\\
    \samplequery& \texttt{?x}   &\leftarrow&   \texttt{?x (actedIn/-actedIn)+ Kevin\_Bacon} &  \uncheckbox{}&\checkbox{}&\uncheckbox{}&\uncheckbox{}&\uncheckbox{}&\uncheckbox{}\\
    \samplequery& \texttt{?area}   &\leftarrow& \texttt{wce -type/(IsL+/dw|dw) ?area} &  \uncheckbox{}&\uncheckbox{}&\checkbox{}&\checkbox{}&\checkbox{}&\uncheckbox{}\\
    \samplequery& \texttt{?person}   &\leftarrow&   \texttt{?person isMarriedTo+/owns/IsL+|owns/IsL+ USA} &  \uncheckbox{}&\checkbox{}&\uncheckbox{}&\checkbox{}&\checkbox{}&\uncheckbox{}\\
    \samplequery& \texttt{?x,?y}   &\leftarrow&  \texttt{?x IsL+/dw+ ?y} &  \uncheckbox{}&\uncheckbox{}&\uncheckbox{}&\uncheckbox{}&\uncheckbox{}&\checkbox{}\\
    \samplequery& \texttt{?x,?y}   &\leftarrow&   \texttt{?x (IsL|dw|rdfs:subClassOf|isConnectedTo)+ ?y} &  \checkbox{}&\uncheckbox{}&\uncheckbox{}&\uncheckbox{}&\uncheckbox{}&\uncheckbox{}\\ 
    \samplequery& \texttt{?x}   &\leftarrow&   \texttt{?x (isConnectedTo/-isConnectedTo)+ S\_Airport} &  \uncheckbox{}&\checkbox{}&\uncheckbox{}&\uncheckbox{}&\uncheckbox{}&\uncheckbox{}\\
    \samplequery& \texttt{?person}   &\leftarrow&  \texttt{?person (wasBornIn/IsL/-wasBornIn)+ JLT}  &  \uncheckbox{}&\checkbox{}&\uncheckbox{}&\uncheckbox{}&\uncheckbox{}&\uncheckbox{}\\
    \samplequery& \texttt{?x}   &\leftarrow&  \texttt{Jay\_Kappraff (livesIn/IsL/-livesIn)+ ?x} &  \uncheckbox{}&\uncheckbox{}&\checkbox{}&\uncheckbox{}&\uncheckbox{}&\uncheckbox{}\\
    \samplequery& \texttt{?x,?y}    &\leftarrow&  \texttt{?x (actedIn/-actedIn)+/hasChild+ ?y}    &  \uncheckbox{}&\uncheckbox{}&\uncheckbox{}&\uncheckbox{}&\uncheckbox{}&\checkbox{}\\ 
    \samplequery& \texttt{?x,?y}  &\leftarrow&  \texttt{?x (wasBornIn/IsL/-wasBornIn)+/isMarriedTo ?y} &  \uncheckbox{}&\uncheckbox{}&\uncheckbox{}&\checkbox{}&\uncheckbox{}&\uncheckbox{}\\
    \samplequery& \texttt{?x,?y}  &\leftarrow&  \texttt{?x (actedIn/-actedIn)+/influences ?y} &  \uncheckbox{}&\uncheckbox{}&\uncheckbox{}&\checkbox{}&\uncheckbox{}&\uncheckbox{}\\
     \samplequery& \texttt{?x}  &\leftarrow&  \texttt{Marie\_Curie (hWP/-hWP)+ ?x} &  \uncheckbox{}&\uncheckbox{}&\checkbox{}&\uncheckbox{}&\uncheckbox{}&\uncheckbox{}\\
     \samplequery& \texttt{?x}        &\leftarrow&  \texttt{London -wasBornIn/(playsFor/-playsFor)+ ?x}  &  \uncheckbox{}&\uncheckbox{}&\checkbox{}&\uncheckbox{}&\checkbox{}&\uncheckbox{}\\
     \samplequery& \texttt{?x}     &\leftarrow&  \texttt{London (-wasBornIn/hWP/-hWP/wasBornIn)+ ?x}  &  \uncheckbox{}&\uncheckbox{}&\checkbox{}&\uncheckbox{}&\uncheckbox{}&\uncheckbox{}\\
     \samplequery& \texttt{?x,?y}        &\leftarrow&  \texttt{?x -actedIn/(-created/influences/created)+ ?y} &  \uncheckbox{}&\uncheckbox{}&\uncheckbox{}&\uncheckbox{}&\checkbox{}&\uncheckbox{}\\
     \samplequery& \texttt{?x,?y}        &\leftarrow&  \texttt{?x -isLeaderOf/(livesIn/-livesIn)+ ?y} &  \uncheckbox{}&\uncheckbox{}&\uncheckbox{}&\uncheckbox{}&\checkbox{}&\uncheckbox{} \\
     \samplequery& \texttt{?x,?y} &\leftarrow&   \texttt{?x (-created/created)+/directed ?y} &\uncheckbox{}&\uncheckbox{}&\uncheckbox{}&\checkbox{}&\uncheckbox{}&\uncheckbox{} \\
     \samplequery& \texttt{?x} &\leftarrow&  \texttt{Lionel\_Messi (playsFor/-playsFor)+/isAff  ?y}  &\uncheckbox{}&\uncheckbox{}&\checkbox{}&\checkbox{}&\uncheckbox{}&\uncheckbox{} \\
     \samplequery& \texttt{?x} &\leftarrow& \texttt{SH (haa|influences)+/(isMarriedTo|hasChild)+ ?x} &\uncheckbox{}&\uncheckbox{}&\checkbox{}&\uncheckbox{}&\uncheckbox{}&\checkbox{} \\
     \samplequery& \texttt{?x,?y} &\leftarrow& \texttt{?x isConnectedTo+/IsL+/dw+/owns+ ?y} &\uncheckbox{}&\uncheckbox{}&\uncheckbox{}&\uncheckbox{}&\uncheckbox{}&\checkbox{} \\
     \samplequery& \texttt{?x,?y} &\leftarrow& \texttt{?x haa/hasChild/(hWP/-hWP)+ ?y} &\uncheckbox{}&\uncheckbox{}&\uncheckbox{}&\uncheckbox{}&\checkbox{}&\uncheckbox{} \\ \hline
  \end{array}$
    \end{tiny}
\end{center}
  \caption[\yago{} queries.]%
  {Queries for the \yago{} dataset\footnotemark.}
 \label{fig:yagoQueries} 
   \vspace{-0.3cm}
    \end{figure}

    \footnotetext{ ``\texttt{isL}'' stands for ``\texttt{IsLocatedIn}'', ``\texttt{dw}'' for ``\texttt{dealsWith}'', ``\texttt{haa}'' for ``\texttt{hasAcademicAdvisor}'', ``\texttt{JLT}'' for ``\texttt{John\_Lawrence\_Toole}'', ``\texttt{hWP}'' for ``\texttt{hasWonPrize}'', ``\texttt{SH}'' for ``\texttt{Stephen\_Hawking}'', ``\texttt{isAff}'' for ``\texttt{isAffiliatedTo}'', ``\texttt{S\_Airport}'' for ``\texttt{Shannon\_Airport}'', and ``\texttt{wce}'' for ``\texttt{wikicat\_Capitals\_in\_Europe}''.}

\paragraph{Concatenated closures} We consider queries of the form $a_1$+/$a_2$+/.../$a_n$+ where $2\leq n \leq 10$. These queries all belong to class $\querycat{6}$.

\paragraph{Non regular queries}

We also consider queries that contain non-regular forms of recursion. These queries are exclusively expressible as \mus{} terms, not as UCRPQs. All of these queries belong to $\querycat{7}$:
\begin{packeditemize}
\item \anbn{} queries: they return the pairs of nodes connected by a path composed of a number of edges labeled $a$ followed by the same number of edges labeled $b$. They are expressed with the following \mus{} term: 
\begin{small}
\begin{flalign*}
&\fixpt{\drop{m}{\rename{trg}{m}{\filt[pred=a]{R}} \NJoin \rename{src}{m}{\filt[pred=b]{R}}}\\
&~~~~~~~~~\cup \drop{m}{\drop{n}{\rename{trg}{m}{\filt[pred=a]{R}} \NJoin \rename{trg}{n}{\rename{src}{m}{X}}\\
&~~~~~~~~~~~~~~\NJoin \rename{src}{n}{\filt[pred=b]{R}}
}}}
\end{flalign*}
\end{small}
\item Same Generation (SG) queries: they return the pairs of nodes that are of the same generation in a graph. 
We use the following term to express them: 
\begin{small}
\begin{flalign*}
    T_{SG}=&\fixpt{\drop{m}{\rename{src}{m}{R} \NJoin \rename{src}{m}{R}}\\
    &~~~~\cup \drop{m}{\drop{n}{\rename{src}{m}{R} \NJoin \rename{trg}{n}{\rename{src}{m}{X}}
    \NJoin \rename{src}{n}{R}
    }}}
\end{flalign*}
\end{small}

\item Filtered SG queries: they compute pairs of nodes in the same generation for a particular predicate $p$ in a graph.
\begin{small}
\begin{flalign*}
    \filt[pred=p]{T_{SG}}
\end{flalign*}
\end{small}
\item Joined SG: they return the pairs of nodes that are of the same generation for a particular set of predicates $P$ in a graph. $P$ is a one column ($pred$) relation that gets joined with the $T_{SG}$ term on the column $pred$:
\begin{small}
\begin{flalign*}
    P \Join T_{SG}
\end{flalign*}
\end{small}
\end{packeditemize}

\paragraph{Uniprot queries} For the synthetic Uniprot datasets, we use the UCRPQ queries shown in Fig.~\ref{fig:uniprotQueries2}.

\begin{figure}[h]
    \begin{center}
\begin{tiny}
    \setttsize{\tiny}
$\begin{array}{|@{\hskip 1pt}l@{\hskip 1pt}|l@{\hskip 1pt}l@{\hskip 1pt}l|@{\hskip 1pt}l@{\hskip 1pt}l@{\hskip 1pt}l@{\hskip 1pt}l@{\hskip 1pt}l@{\hskip 1pt}l@{\hskip 1pt}|} \hline 
    \query{id}&Query&&&\querycat{1}&\querycat{2}&\querycat{3}&\querycat{4}&\querycat{5}&\querycat{6}\\ \hline \rule{0pt}{3ex}  
    \samplequery& \texttt{?x,?y} &\leftarrow& \texttt{?x -hKw/(ref/-ref)+ ?y } &\uncheckbox{}&\uncheckbox{}&\uncheckbox{}&\uncheckbox{}&\checkbox{}&\uncheckbox{} \\
\samplequery& \texttt{?x,?y} &\leftarrow& \texttt{?x -hKw/(enc/-enc)+ ?y } &\uncheckbox{}&\uncheckbox{}&\uncheckbox{}&\uncheckbox{}&\checkbox{}&\uncheckbox{} \\
\samplequery& \texttt{?x} &\leftarrow& \texttt{C (occ/-occ)+ ?x} &\uncheckbox{}&\uncheckbox{}&\checkbox{}&\uncheckbox{}&\uncheckbox{}&\uncheckbox{} \\
\samplequery& \texttt{?x,?y} &\leftarrow& \texttt{?x int+/(occ/-occ)+/(hKw/-hKw)+ ?y} &\uncheckbox{}&\uncheckbox{}&\uncheckbox{}&\uncheckbox{}&\uncheckbox{}&\checkbox{} \\
\samplequery& \texttt{?x} &\leftarrow& \texttt{?x (enc/-enc | occ/-occ)+ C} &\uncheckbox{}&\checkbox{}&\uncheckbox{}&\uncheckbox{}&\uncheckbox{}&\uncheckbox{} \\
\samplequery& \texttt{?x,?y} &\leftarrow& \texttt{?x int+/(occ/-occ)+ ?y} &\uncheckbox{}&\uncheckbox{}&\uncheckbox{}&\uncheckbox{}&\uncheckbox{}&\checkbox{} \\
\samplequery& \texttt{?x,?y} &\leftarrow& \texttt{?x int+/(enc/-enc)+ ?y} &\uncheckbox{}&\uncheckbox{}&\uncheckbox{}&\uncheckbox{}&\uncheckbox{}&\checkbox{} \\
\samplequery& \texttt{?x,?y} &\leftarrow& \texttt{?x int/(enc/-enc)+ ?y } &\uncheckbox{}&\uncheckbox{}&\uncheckbox{}&\uncheckbox{}&\checkbox{}&\uncheckbox{} \\
\samplequery& \texttt{?x,?y} &\leftarrow& \texttt{?x -hKw/int/ref/(auth/-auth)+ ?y} &\uncheckbox{}&\uncheckbox{}&\uncheckbox{}&\uncheckbox{}&\checkbox{}&\uncheckbox{} \\
\samplequery& \texttt{?x,?y} &\leftarrow& \texttt{?x (enc/-enc)+/hKw ?y} &\uncheckbox{}&\uncheckbox{}&\uncheckbox{}&\checkbox{}&\uncheckbox{}&\uncheckbox{} \\
\samplequery& \texttt{?x} &\leftarrow& \texttt{?x (enc/-enc)+ C} &\uncheckbox{}&\checkbox{}&\uncheckbox{}&\uncheckbox{}&\uncheckbox{}&\uncheckbox{} \\
\samplequery& \texttt{?x,?y,?z,?t} &\leftarrow& \texttt{?x (enc/-enc)+ ?y, ?x int+ ?z, ?x ref  ?t} &\uncheckbox{}&\uncheckbox{}&\uncheckbox{}&\uncheckbox{}&\checkbox{}&\checkbox{} \\
\samplequery& \texttt{?x,?y} &\leftarrow& \texttt{?x (int|(enc/-enc))+ ?y, C (occ/-occ)+ ?y} &\uncheckbox{}&\uncheckbox{}&\checkbox{}&\uncheckbox{}&\uncheckbox{}&\checkbox{} \\
\samplequery& \texttt{?x} &\leftarrow& \texttt{?x int+/ref ?y, C (auth/-auth)+ ?y} &\uncheckbox{}&\uncheckbox{}&\checkbox{}&\checkbox{}&\uncheckbox{}&\uncheckbox{} \\
\samplequery& \texttt{?x} &\leftarrow& \texttt{?x int+/ref ?y, C -pub/(auth/-auth)+ ?y}   &\uncheckbox{}&\uncheckbox{}&\checkbox{}&\checkbox{}&\checkbox{}&\uncheckbox{} \\
\samplequery& \texttt{?x} &\leftarrow& \texttt{ C -pub/(auth/-auth)+ ?x}  &\uncheckbox{}&\uncheckbox{}&\checkbox{}&\uncheckbox{}&\checkbox{}&\uncheckbox{} \\
\samplequery& \texttt{?x,?y} &\leftarrow& \texttt{?x -occ/int+/occ ?y} &\uncheckbox{}&\uncheckbox{}&\uncheckbox{}&\checkbox{}&\checkbox{}&\uncheckbox{} \\
\samplequery& \texttt{?x,?y} &\leftarrow& \texttt{?x (-ref/ref)+ ?y} &\checkbox{}&\uncheckbox{}&\uncheckbox{}&\uncheckbox{}&\uncheckbox{}&\uncheckbox{} \\
\samplequery& \texttt{?x,?y} &\leftarrow& \texttt{?x int/ref/(-ref/ref)+ ?y} &\uncheckbox{}&\uncheckbox{}&\uncheckbox{}&\uncheckbox{}&\checkbox{}&\uncheckbox{} \\
\samplequery& \texttt{?x} &\leftarrow& \texttt{C (ref/-ref)+ ?x} &\uncheckbox{}&\uncheckbox{}&\checkbox{}&\uncheckbox{}&\uncheckbox{}&\uncheckbox{} \\
\samplequery& \texttt{?x,?y} &\leftarrow& \texttt{?x (-ref/ref)+/(auth|pub) ?y} &\uncheckbox{}&\uncheckbox{}&\uncheckbox{}&\checkbox{}&\uncheckbox{}&\uncheckbox{} \\
\samplequery& \texttt{?x,?y} &\leftarrow& \texttt{?x int/(occ/-occ)+ ?y} &\uncheckbox{}&\uncheckbox{}&\uncheckbox{}&\uncheckbox{}&\checkbox{}&\uncheckbox{} \\
\samplequery& \texttt{?x} &\leftarrow& \texttt{C int/(enc/-enc|occ/-occ)+ ?x} &\uncheckbox{}&\uncheckbox{}&\checkbox{}&\uncheckbox{}&\checkbox{}&\uncheckbox{} \\
\samplequery& \texttt{?x} &\leftarrow& \texttt{C (enc/-enc)+ ?x} &\uncheckbox{}&\uncheckbox{}&\checkbox{}&\uncheckbox{}&\uncheckbox{}&\uncheckbox{} \\
\samplequery& \texttt{?x,?y} &\leftarrow& \texttt{?x -hKw/(occ/-occ)+ ?y } &\uncheckbox{}&\uncheckbox{}&\uncheckbox{}&\uncheckbox{}&\checkbox{}&\uncheckbox{} \\ \hline 
\end{array}$
\end{tiny}
\end{center}
\caption[Uniprot queries.]%
{Uniprot queries\footnotemark.}
\label{fig:uniprotQueries2} 
\vspace{-0.3cm}
\end{figure}

\footnotetext{ ``\texttt{int}'' stands for ``\texttt{interacts}'', ``\texttt{enc}'' for ``\texttt{encodes}'', ``\texttt{occ}'' for ``\texttt{occurs}'', ``\texttt{hKw}'' for ``\texttt{hasKeyword}'', ``\texttt{ref}'' for ``\texttt{reference}'', ``\texttt{auth}'' for ``\texttt{authoredBy}'', and ``\texttt{pub}'' for ``\texttt{publishes}''.}

\subsection{Results}
We report on experimental results and analyse them. We measure the time spent in evaluating queries by the different systems, in seconds.
For each set of experiments, we define a timeout value. Whenever the time spent in evaluating a query reaches this timeout value, we consider that the query evaluation did not terminate within a reasonable time.
On charts, the timeout value corresponds to the maximum value on the y-axis. 
Some systems crashed in some query evaluations. In charts, this is denoted by the presence of a red cross on a time bar. 
The observed crashes are out of memory and timeout failures. They are due to the amount of data processed and transferred over the network that exceeds the capacity of the machines. This amount of data is linked to the size of intermediate results produced by the query evaluation.
The other cases correspond to query evaluations where the system answered correctly. The plotted times represent the average running times over three executions.

\subsubsection{\distmuir{} recursive plans evaluation}\label{sec:xprecplans} 

\begin{figure*}[h]
    \centering
        \includegraphics[width=0.8\textwidth]{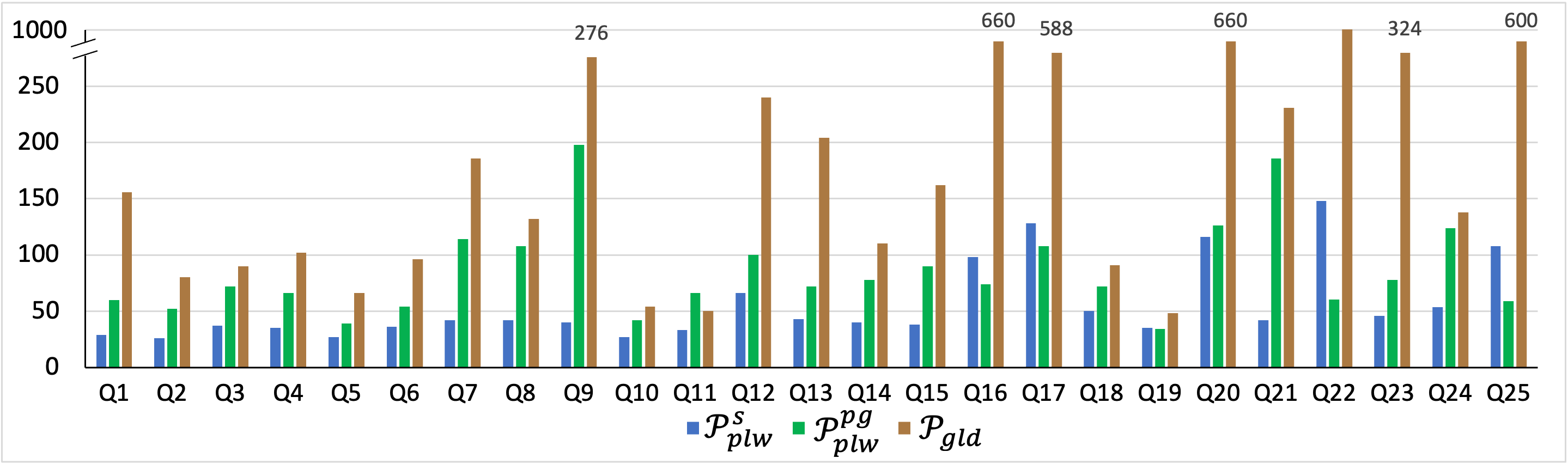}
        \caption{Running times of $\planparallelloop$ and $\plangloballoop$ plans on Yago.}\label{fig:setrddpostgres}
\end{figure*}

Fig.~\ref{fig:setrddpostgres} presents the time spent in evaluating each of the \distmuir{} plans (Sec.~\ref{sec:distribution}) for UCRPQs on the Yago dataset. We observe that the $\planparallelloop$ plans are faster than $\plangloballoop$. This illustrates the interest of the communication cost reduction performed by $\planparallelloop$. As explained in Sec.~\ref{sec:distribution}, $\plangloballoop$ requires communications between the workers at each step of the recursion while $\planparallelloop$ does not.

\subsubsection{UCRPQs on Yago: comparison with other systems}~\label{sec:xpresyago}

\begin{figure*}[h]
    \centering
    \includegraphics[width=0.85\textwidth]{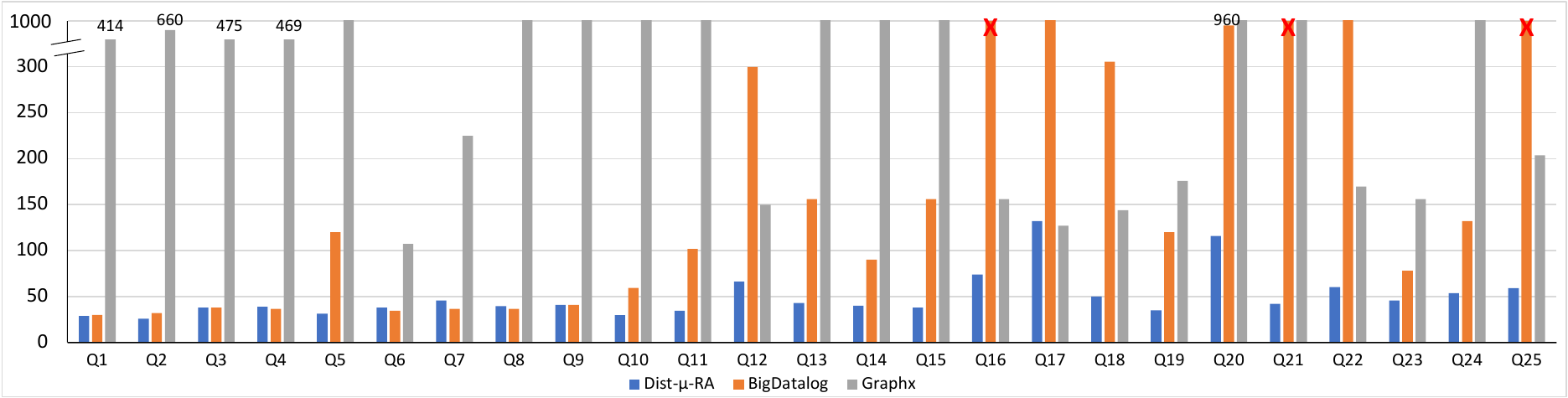}
    \caption{Running times on Yago. A timeout is set at 1,000 s.}\label{fig:ucrpqyagoexec}
\end{figure*}

Fig.~\ref{fig:ucrpqyagoexec} shows the performance results of \distmuir{}, BigDatalog and GraphX for queries $\query{1}-\query{25}$ on the Yago dataset.  

First, these results show that \distmuir{} is much faster than GraphX overall. 
We attribute GraphX's lower performance to its Pregel model, where each node must track its ancestors that satisfy a given regular path query (or part of it) and transmit this information to successors to find pairs of nodes that satisfy the entire query. While GraphX is efficient for many graph algorithms \cite{GonzalezXDCFS14}, it may not be ideal for this type of query. The only instance where GraphX matches \distmuir{}'s performance is for $\query{17}$, where filtering is done early in the query (see Sec.~\ref{sec:xpsystems}).

Second, these results show that \distmuir{} provides much faster performance than Bigdatalog for all the classes  $\querycat{2}$-$\querycat{6}$, and comparable performance for class  $\querycat{1}$.

One explanation for the difference in performance of $\query{5}$ of class $\querycat{2}$ is that it requires reversing a fixpoint term first before pushing the filter ``Kevin Bacon''. This fixpoint reversal is not supported by Datalog's Magic Sets optimization technique (see Sec.~\ref{sec:relatedworks} for more details). 
Another example is $\query{24}$ of class $\querycat{6}$ where \distmuir{} merges fixpoint terms (which BigDatalog is unable to do).  
Overall, the optimizations in \distmuir{} are more effective. We noticed that this is particularly true when the size of intermediate results is large ($\query{5}$ and $\query{10}-\query{25}$).

\subsubsection{Concatenated closures}
We now evaluate concatenated closure queries (which belong to $\querycat{6}$) on the graph obtained from \rndfile{0.001}{100k}. The graph edges are randomly labeled from a set of 10 different labels. Results are shown in Fig.~\ref{fig:concatenatedclosures}.  \distmuir{} is faster on all queries. The time difference between \distmuir{} and BigDatalog
for a query with $n$ concatenations ($a_1$+/.../$a_n$+)
becomes larger when $n$ increases. BigDatalog fails for queries where $n \geq 5$ and GraphX crashes on all queries.
The plans that are selected in \distmuir{} for the execution of these queries apply a mixture of the rewritings that ``push joins'' and ``merge fixpoints'' (see Sec.~\ref{sec:architecture}). These results also indicate that optimizations introduced by these rewritings provide significant performance gains for class $\querycat{6}$.

\subsubsection{Non regular queries}
\begin{figure*}[h]
\begin{minipage}{0.79\textwidth}
        \includegraphics[width=1\textwidth]{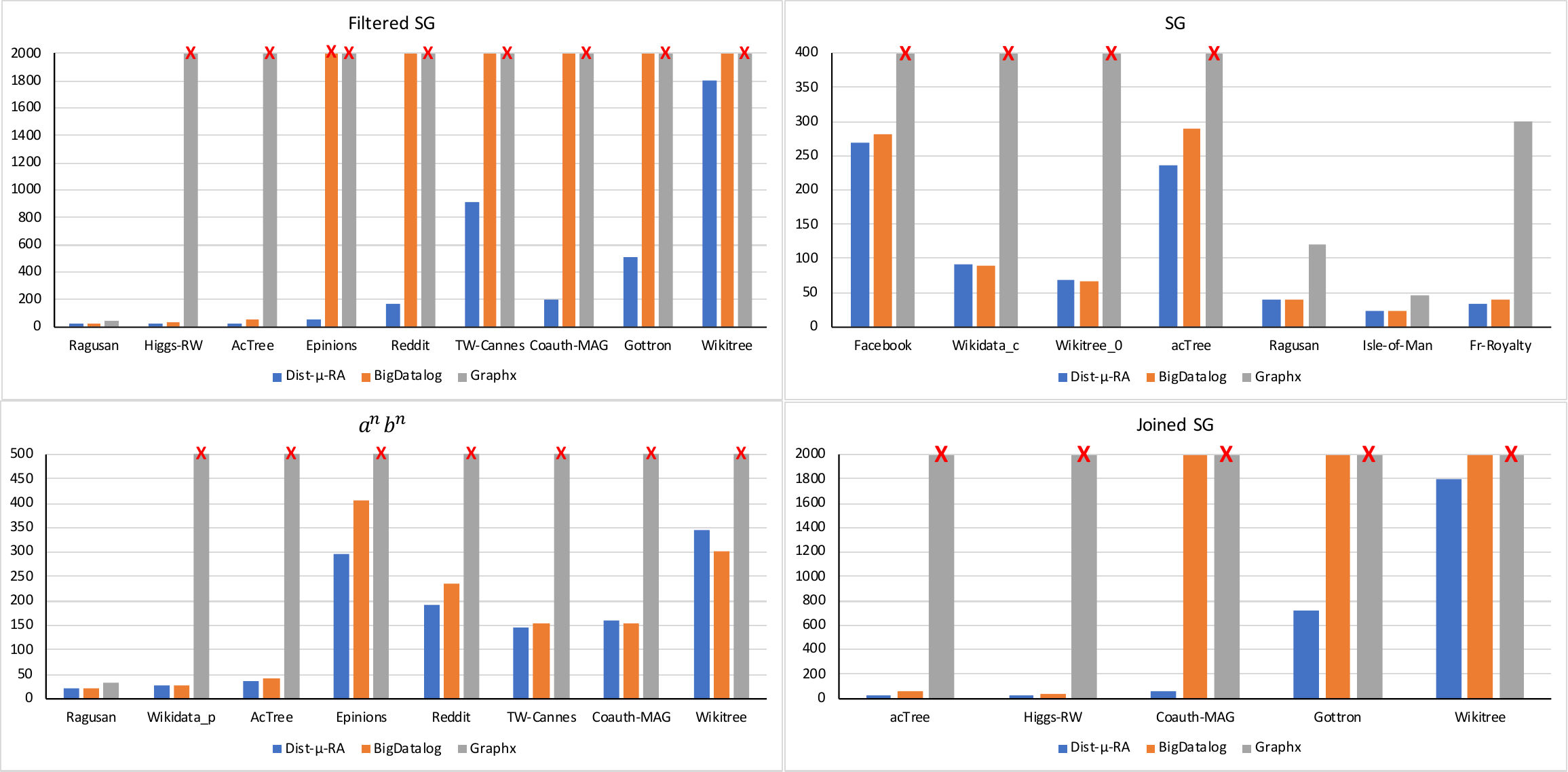}
    \caption{\mus{} queries running times. A timeout is set to 2000s.}\label{fig:muexec}
\end{minipage}
    \begin{minipage}{0.2\textwidth}
        \centering
        \includegraphics[width=\textwidth]{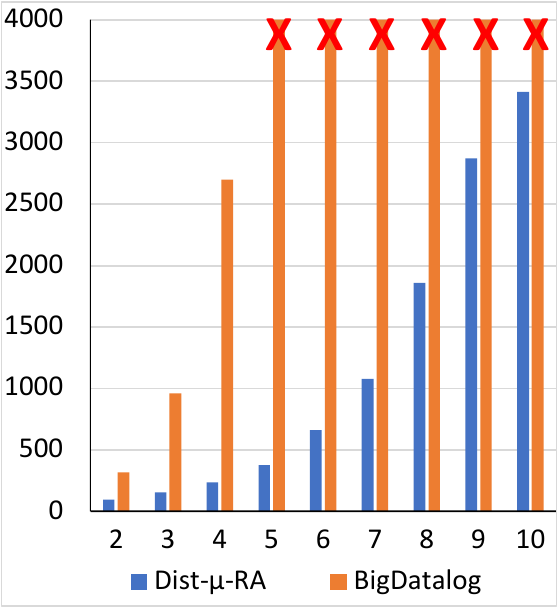}
        \caption{Running times of concatenated closures queries.}\label{fig:concatenatedclosures}
    \end{minipage}

\end{figure*}

The execution times for these queries of class $\querycat{7}$ are given in Fig.~\ref{fig:muexec}.
On the basic SG and \anbn{} queries, \distmuir{} and BigDatalog have comparable execution times.  \distmuir{} is faster on Filtered SG and Joined SG queries.

\subsubsection{UCRPQs on Uniprot}

\begin{figure*}[h]
    \centering
    \includegraphics[width=0.9\textwidth]{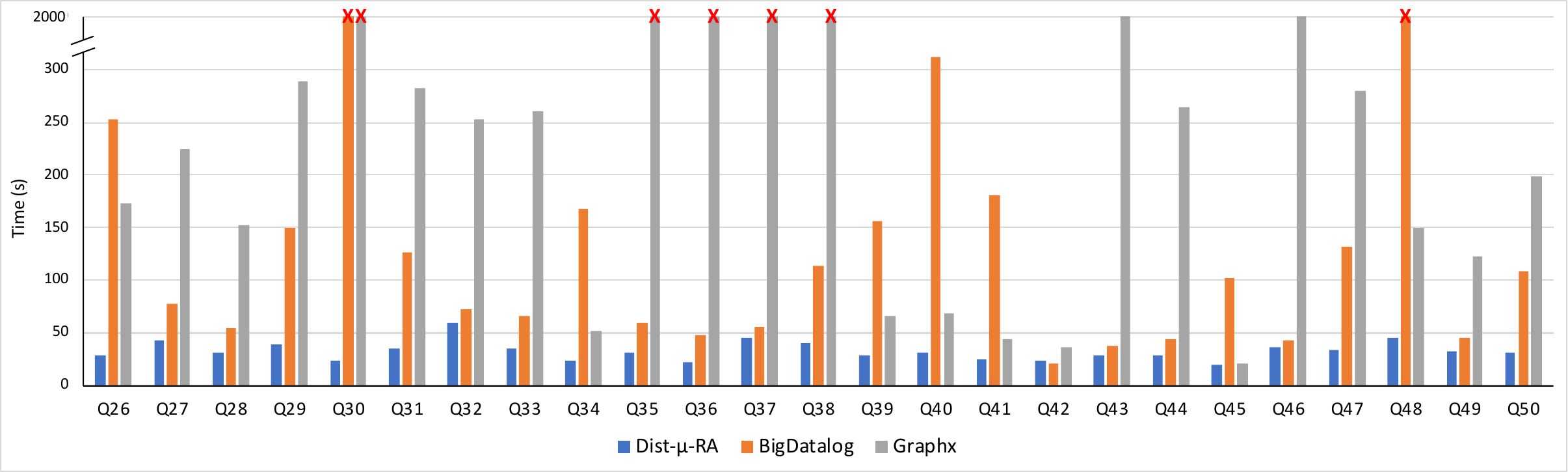}
    \caption{Running times on \uniprot{1M}. A timeout is set to 2,000 s.}\label{fig:ucrpqexec}
\end{figure*}

The results reported in Fig~\ref{fig:ucrpqexec} show that \distmuir{} is the only system that answers all of the queries. Furthermore, \distmuir{} is faster on all queries belonging to $\querycat{2-6}$ (except $\query{42}$ where the size of the transitive closure is small). 
 
\paragraph{Further scalability banchmarks on Uniprot}
\begin{figure*}[h]
    \centering
    \includegraphics[width=0.9\textwidth]{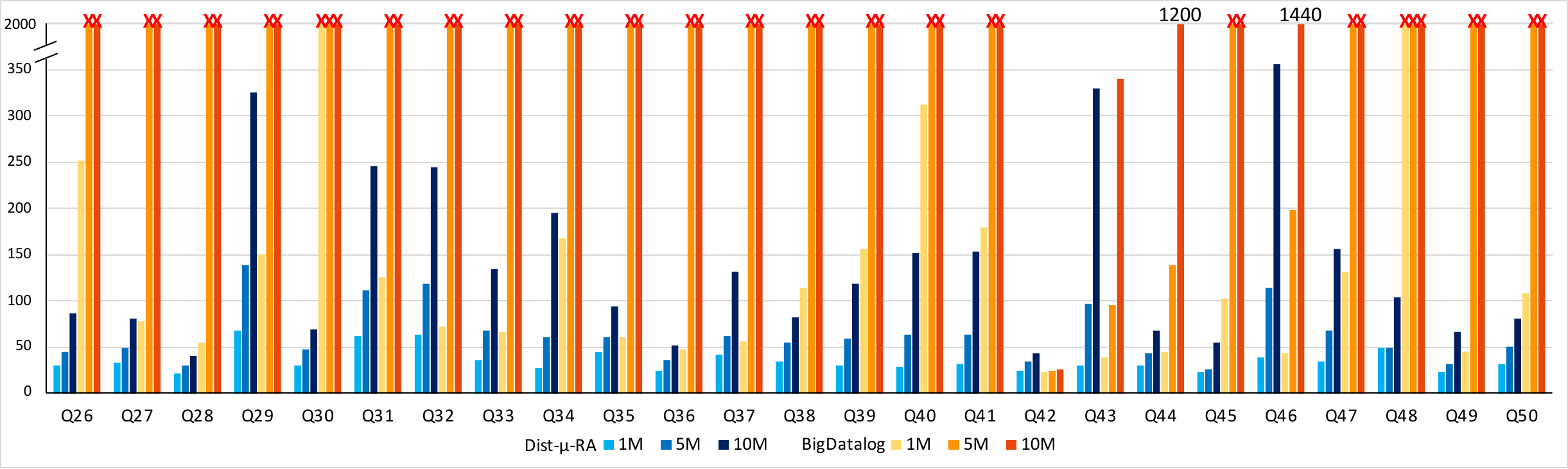}\vspace{-0.2cm}\caption{\distmuir{} and BigDatalog running times on Uniprot graphs of different sizes}\label{fig:murabdlscalability}
\end{figure*}

We report on further scalability benchmarks where \distmuir{} and BigDatalog execution times are compared for each Uniprot query on generated \uniprot{n} graphs with varying sizes of 1M, 5M and 10M edges.  Results are shown in Fig.~\ref{fig:murabdlscalability}.
Results indicate that BigDatalog fails in 44 cases out of 75 query evaluations. \distmuir{} answers all of them and scales better.

Notice that, for comprehensive benchmarking, queries and graph sizes have been selected so as to cover a wide range of result sizes. $\query{40}$ is one of the queries with the smallest result size (14K records for \uniprot{10M}) and $\query{46}$ is one with the largest (around 1.5B records for \uniprot{10M}, which is 150 times the size of the graph).

\subsubsection{Cost model evaluation}
\begin{figure}
    \centering
\includegraphics[width=0.4\textwidth, keepaspectratio]{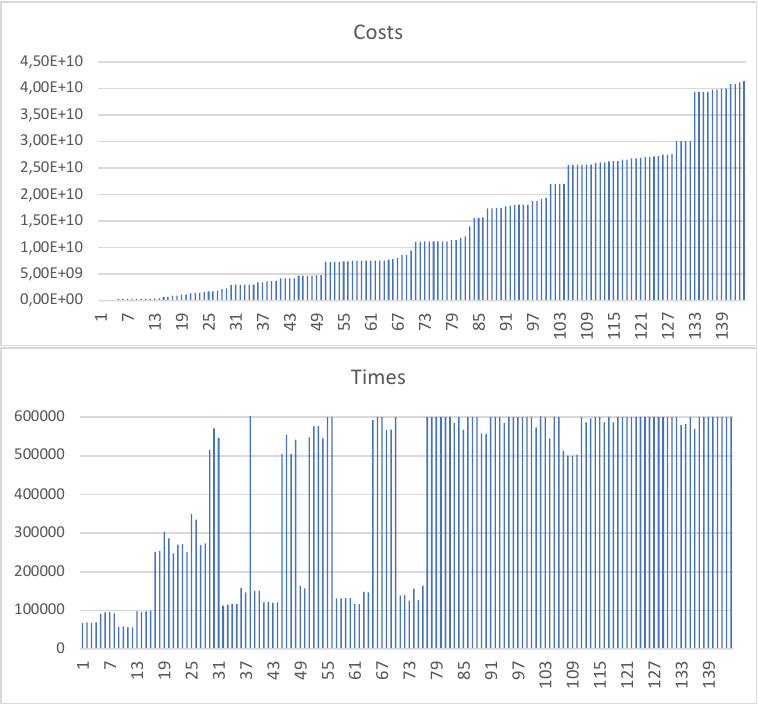}
\caption{Estimated costs and corresponding evaluation times of all equivalent plans of query $\query{24}$.}\label{fig:costmodel}
\end{figure}

The cost model we used for the reported experiments is based on the one introduced in \cite{muideen-cikm2020}, which has been shown to be slightly more accurate for recursive queries than the one implemented in PostgreSQL \cite{muideen-cikm2020}. 
To evaluate the cost model used for selecting the best plan, we have computed on each of the 16 Yago queries, both the estimated costs and the execution times of all equivalent terms of the query\footnote{The detailed results can be found in \cite{distmura-page}}. 
For instance, Fig.~\ref{fig:costmodel} shows estimated costs ranked in increasing order (top chart) with their corresponding evaluation time (bottom chart) for the Yago query $\query{24}$. 
More generally, we observe that, on average, the evaluation time of the selected term is within the top 14.7\% execution times, the selected term is 58\% faster than the average evaluation time of all equivalent plans (timeout set to 600s) and 20\% slower than the term having the best evaluation time. 
Our approach remains modular and could benefit from further advances in cost estimations.

\subsection{Summary}

Overall, for all query classes, \distmuir{} is significantly more efficient compared to GraphX. For query classes $\querycat{2-6}$ and some queries in $\querycat{7}$, \distmuir{} is more efficient than BigDatalog, especially for large intermediate query result sizes.
For query class $\querycat{1}$ and some queries in $\querycat{7}$, \distmuir{} and BigDatalog have a comparable performance. Our empirical findings tend to indicate that for these cases the various optimizations techniques of \distmuir{} and Bigdatalog have limited impact.

\section{Related Works}\label{sec:relatedworks}
 
In order to evaluate expressive queries (such as UCRPQs) over graphs, it is essential for a system to be able to: (i) support recursion and the optimization of recursive terms, and (ii) provide distribution of both data and computations. We examine and compare to the closest related works along these two aspects below.

For (i) we have choosen to build our system using \mus{} \cite{mura-sigmod20} since it offers more optimization opportunities without sacrifying expressivity, in particular with respect to approaches based on Datalog and RA~\cite{mura-sigmod20}. 
The Datalog research line \cite{datalog-1, datalog-2, datalog-3, datalog-4, datalog-5, bigdatalog, datalog-7, myria2015} has developed various methods for optimizing recursive queries expressed in Datalog, including magic-sets~\cite{datalog-magic-sets, datalog-magic-sets-2, datalog-opt}, demand transformations~\cite{demand-driven-datalog-2011}, automated reversals~\cite{right-left-linear-datalog}, and the FGH rule~\cite{fgh-rule-datalog-2022}. %
Although Datalog has a syntax that significantly differs from relational algebra (RA), the effects of magic-sets~\cite{datalog-magic-sets, datalog-magic-sets-2, datalog-opt} and demand transformations~\cite{demand-driven-datalog-2011} can be seen as equivalent to pushing specific types of selections and projections. These optimization techniques are highly sensitive to whether the Datalog program is written in a left-linear or right-linear form, but automated reversal techniques~\cite{right-left-linear-datalog} can be used to fully leverage them. The framework introduced in~\cite{fgh-rule-datalog-2022} unifies magic-sets with semi-naïve evaluation and introduces the FGH rule to optimize recursive Datalog programs involving aggregation. %
Datalog engines do not explore alternative execution plans but instead rely on heuristics to determine an efficient evaluation strategy. However, regardless of which combination of existing Datalog optimizations a given engine employs, it currently lacks the capability to merge recursions in the same way as the $\mu$-RA approach~\cite{mura-sigmod20}. This limitation arises because, in a Datalog program that represents the optimized translation of A$^+$/B$^+$, at least one of the two transitive closures, A$^+$ or B$^+$, will always be fully materialized—even when A$^+$/B$^+$ has no solution. On real-world datasets, this can result in Datalog query evaluation being an order of magnitude slower than query execution in RA-based systems, as observed in~\cite{mura-sigmod20}. %
Since $\mu$-RA already enables a richer set of execution plans than Datalog, Dist-$\mu$-RA expands this space even further. By incorporating distributed execution strategies and partitioning techniques, it explores additional evaluation plans beyond the capabilities of any Datalog engine.

\label{sec:rwformalisms}

Concerning the distribution aspect (ii), the seminal systems MapMeduce and Dryad are known to be inefficient for iterative applications~\cite{CarboneKEMHT15}.
     Spark~\cite{ZahariaXWDADMRV16} and Flink~\cite{CarboneKEMHT15} were introduced to improve upon these systems and became prevalent for large scale and data-parallel computations. 
    Work in~\cite{Katsogridakis2017} proposes a technique that improves Spark task scheduling and thus performance for iterative applications. This system-level optimization is transparent for Spark applications and thus \distmuir{} can directly benefit from it.

Systems specifically designed for large-scale graph processing include Google's Pregel~\cite{googlepregel2010}, Giraph~\cite{giraph} an open-source system based on the Pregel model, GraphLab~\cite{graphlab2012} and Powergraph~\cite{powergraph2012}. GraphX~\cite{GonzalezXDCFS14} is a Spark library for graph processing that offers a Pregel API to perform recursive computations. 
Pregel is based on the Bulk Synchronous Parallel model. A Pregel program is composed of supersteps. At each superstep, a vertex receives messages sent by other vertices at the previous iteration and processes them to update its state and send new messages. Computation stops when no new message is sent. Is is not straightforward to evaluate UCRPQs in Pregel. An automata like algorithm needs to be written to know which stage of the regular query each processed path has reached. The idea is to traverse the paths in the graph (by sending messages from vertices to their neighbors) while traversing the regular query. 
    \cite{wang2020} proposes a system that implements RPQ queries on GraphX and proposes optimizations to reduce communications between nodes. 
    In all these systems, selections can be pushed in one direction only. For instance, if the program traverses the regular query from left to right, the execution of the program naturally computes the filters and edge selections occuring before a recursion  first, thereby pushing these operations in the recursion. Selections which occur after the recursion cannot be pushed.
    Additionally, communications between workers happen in every recursion superstep, which is avoided by the $\planparallelloop$ plan in \distmuir{}.

Distributed systems with higher-level query language support have been developed.
The Spark SQL~\cite{sparksql2015} library enables the user to write SQL queries and process relational data using \dataset{}s or \dataframe{}s.  However, recursion is not supported.  DryadLINQ~\cite{dryadlink2008} that exposes a declarative query language on top of Dryad (or Pig Latin~\cite{piglatin2008} on top of Hadoop MapReduce) has the same limitation. 
TitanDB~\cite{titandb} is a distributed graph database that supports the Gremlin query language.  
Gremlin provides primitives for expressing graph traversals. It is able to express UCRPQ queries with its own syntax. However, these systems do not provide optimization techniques comparable to the ones we propose.

SociaLite~\cite{socialite2013} is an extension of Datalog for social network graph analysis. 
Its distributed implementation runs queries on a cluster of multi-core machines in which workers communicate using message passing. SociaLite does not offer a distribution plan equivalent to $\planparallelloop$ where recursion can be executed without communication between workers at every step.

Dataflow systems like Naiad~\cite{naiad} and Banyan~\cite{banyan} introduce optimizations to minimize synchronization and coordination overhead in iterative computations. However, they do not address automated partitioning optimization or cross-node data transfer reduction such as those explored in this paper. Dist-$\mu$-RA, built on relational algebra with recursion, operates at a higher level and can use these systems as backends. Unlike Naiad and Banyan, which rely on fixed traversal directions, Dist-$\mu$-RA supports flexible query transformations like fixpoint merging and reversal, leading to significant performance gains.   
Myria~\cite{myria2015} is a distributed dataflow system which supports a subset of Datalog. Myria was shown more efficient than Naiad in previous work \cite{myria2015}. Queries are translated into query plans executed on a parallel relational engine. Myria supports incremental evaluation of recursion and provides synchronous and asynchronous modes. It does not support advanced logical optimizations of the recursive query plan like pushing joins in fixpoints nor merging fixpoints. Myria does not either propose a distribution plan equivalent to $\planparallelloop$. In practice, \distmuir{} significantly outperforms Myria (see experiments in \cite{distmura-page}).

RaSQL~\cite{rasql2019} proposes an extension of SQL with some aggregate operations in recursion. Queries are compiled to Spark SQL to be distributed and executed on Spark. RaSQL does not propose rules to push selections in the fixpoint operator nor to merge fixpoints. RaSQL proposes a decomposable plan for recursion similar to the one in BigDatalog but has no automated technique to distribute data. The RaSQL implementation is not available for benchmarking.

BigDatalog~\cite{bigdatalog} is a recursive Datalog engine that runs on Spark. 
It uses the Datalog GPS technique~\cite{seib1991} that analyses Datalog rules to identify decomposable Datalog programs and determine how to distribute data and computations. These ideas are tied to Datalog and are not applicable to the relational algebra. The present work proposes a new method specifically designed for recursive relational algebraic terms. It uses the \mus{} filter pushing technique to automatically repartition data. Compared to BigDatalog, \distmuir{} is superior because it supports optimizations that BigDatalog, like any Datalog-based engine, is unable to provide, as explained previously. 
In practice, \distmuir{} provides superior performance on a wider range of query classes, as reported by the experiments in Section~\ref{sec:experiments}.

\section{Conclusion}
We propose a method for the evaluation of recursive algebraic terms in  a distributed manner. The method generates independent smaller parallel loops on worker nodes instead of executing a single global loop on the driver node. The advantage of the parallel local loops is a minimization of the amount of data shuffled between worker nodes.  We applied this approach to recursive graph queries on real and synthetic datasets. Experimental results show significant performance gains compared to the state-of-the-art. Gains are due to local parallel loops that reduce communication costs, which in turn results in an improvement of the overall query evaluation time.

\bibliographystyle{IEEEtran}
\bibliography{biblio}
\end{document}